\documentclass[aps,prd,onecolumn,preprintnumbers,groupedaddress,showpacs,nofootinbib,amssymb]{revtex4}
\usepackage{graphicx,color}
\usepackage{amsmath}
\usepackage{amssymb}
\usepackage{amsfonts}
\usepackage{bm}
\usepackage{cancel}

\newcommand{\e}{\mathrm{e}}

\allowdisplaybreaks[4] 
\begin{document}

\preprint{KEK-TH-2745, KEK-Cosmo-0389}
\title{ Dynamical Black Hole in the accelerating Universe approaching the future singularity \\
-- Possible origin of (super-)massive black holes --
}
\author{Shin'ichi~Nojiri,$^{1,2}$}
\email{nojiri@nagoya-u.jp}
\author{Sergei~D.~Odintsov,$^{3,4}$}
\email{odintsov@ieec.cat}
\affiliation{$^{1)}$ Theory Center, High Energy Accelerator Research Organization (KEK), \\
Oho 1-1, Tsukuba, Ibaraki 305-0801, Japan \\
$^{2)}$ Kobayashi-Maskawa Institute for the Origin of Particles
and the Universe, Nagoya University, Nagoya 464-8602, Japan \\
$^{3)}$ ICREA, Passeig Luis Companys, 23, 08010 Barcelona, Spain\\
$^{4)}$ Institute of Space Sciences (IEEC-CSIC) C. Can Magrans
s/n, 08193 Barcelona, Spain}

\begin{abstract}

We construct and investigate the dynamical black hole spacetime embedded in the expanding universe filled with cosmic fluid, such as dark energy.  
When the equation of state (EoS) parameter of the fluid is a constant, we find exact solutions of the Einstein equation where the 
Schwarzschild black hole is embedded in the 
expanding universe. 
This solution differs from the well-known McVittie metric, where the EoS parameter is not a constant but rather depends on the radial coordinate. 
It is shown that a dynamical black hole grows with the expansion of the universe. 
If primordial black holes are created before or during inflation, above dynamical black holes might be the origin of the supermassive black holes 
at the centre of galaxies, massive black holes suggested by the GW231123 event, and also the dark matter. 
The case where the cosmic fluid EoS is more general is also considered so that the universe enters the epoch of finite-time future singularity. 
Thermodynamics and the behaviour of black holes around different future singularities are carefully investigated. 
It is then demonstrated that the black hole horizon enhances the tidal force, but near the horizon, the tidal force works to press the extended object, 
which is in contrast with a massive body near to future singularity. 

We also propose a new type of future singularity where the singularity inside the black hole is a sphere with a finite radius. 
When the radius of the spherical singularity becomes larger than the radius of the black hole horizon, it becomes naked. 
The universe may end up with a cosmic doomsday when the radius of the singularity becomes infinite. 

\end{abstract}

\maketitle

\section{Introduction}\label{SecI}

The elucidation of the mechanism of the accelerated expansion of the universe, discovered at the end 
of the last century \cite{SupernovaSearchTeam:1998fmf, SupernovaCosmologyProject:1998vns}, is probably the greatest challenge in modern century physics. 
The accelerated expansion might be generated by an unknown fluid called dark energy. 
The behaviour of the dark energy is controlled by the effective equation of state parameter $w$, which is the ratio of the pressure $p$ to the energy density $\rho$, 
\begin{align}
\label{EoS0}
p=w \rho\, .
\end{align}
The accelerated expansion can occur if $w<-\frac{1}{3}$, and the cosmological constant case corresponds to $w=-1$. 
The recent observations \cite{Planck:2018vyg,CosmoVerse:2025txj} do not exclude the possibility that $w$ is less than $-1$, $w<-1$, 
and furthermore there is a possibility that the value of $w$ changes dynamically \cite{DESI:2024mwx,DESI:2025zgx}. 
The dark energy with $w<-1$ is called phantom \cite{Caldwell:1999ew, Caldwell:2003vq}. 
If the universe expansion is governed by phantom dark energy, the universe will end with the rapid expansion at an infinite rate, which is called the Big Rip. 
Near the Big Rip singularity, any extended objects with finite size are torn and ripped due to the tidal force. 
There are other types of future singularity like the one of Ref.~\cite{Barrow:2004xh}, which is much weaker than the Big Rip singularity. 
The types of future singularities in cosmology are classified in \cite{Nojiri:2005sx}. 
Different properties of the future singularities, as well as their behaviour at the accelerating universe, have been well-studied in 
Refs.~\cite{Nojiri:2009pf, Nojiri:2006ww, Dabrowski:2009kg, Odintsov:2022umu, Perivolaropoulos:2016nhp, Antoniou:2016obw, LimaJunior:2025uyj, 
Nesseris:2004uj, Elizalde:2004mq, Nojiri:2005sr, Bamba:2008ut, Bouhmadi-Lopez:2006fwq, Bouhmadi-Lopez:2007xco, Dabrowski:2003jm, 
Dabrowski:2004bz, Nojiri:2004ip, Brevik:2024ozg}, for review, see Refs.~\cite{deHaro:2023lbq, Trivedi:2023zlf}.

The purpose of this work is to investigate dynamical time-dependent black hole properties in the accelerating cosmological background approaching the future singularity. 
General spherical symmetric and dynamical black hole solutions have been constructed by using two scalar fields \cite{Nojiri:2020blr}. 
One of the two scalar fields, however, becomes a ghost, but in Ref.~\cite{Nojiri:2023dvf}, it has been shown that some constraints can eliminate the ghost. 
Note that there is a point of view expressed in Ref.~\cite{Gaztanaga:2022fhp} that our universe may represent a kind of dynamical black hole.

In this work, we consider dynamical black hole solutions generated by cosmic fluid. 
First of all, we show that the Schwarzschild black hole embedded in the universe, which is expanding due to the fluid with a constant 
EoS parameter $w$ (\ref{EoS0}) is an exact solution. 
In the solution, the black hole grows up accompanied by the expansion of the universe, that is, the mass and the radius of the horizon become larger with the expansion. 
This is due to the accretion of the fluid, which may be dark energy. 

Recently, in \cite{LIGOScientific:2025rsn}, a gravitational wave event GW231123 was reported. 
The event seems to be consistent with the merger of two black holes with large masses of $137^{+22}_{-17}\, M_\odot$ and $103^{+20}_{-52}\, M_\odot$. 
Here $M_\odot$ is the solar mass. 
The black holes with these mass scales are believed not to be created by the collapse of astrophysical objects.
The black hole solution of this paper shows that a black hole can grow due to the expansion of the universe. 
This could indicate that the masses of the primordial black holes \cite{Carr:1975qj, Carr:2020gox, Sasaki:2018dmp} during inflation increase rapidly, and 
there appear very massive black holes after inflation. 
This scenario could explain the origin of the black holes in the GW231123 event, in addition to the dark matter and also supermassive black holes at the centre of galaxies. 

It is also shown that for a found black hole near to future singularity, the existence of the black hole horizon enhances the tidal force, which generates the rip of the extended object, 
in most regions. 
Near the horizon, however, the tidal force works to press the extended object, potentially crushing it. 

As an exact solution of the Einstein equation, the McVittie metric \cite{McVittie:1933zz} is well-known. 
The EoS parameter for the McVittie solution is not a constant but depends on the radial coordinate, which is different from that in the solution with a constant 
EoS parameter of this paper. 
Furthermore, in the McVittie solution, the horizon radius shrinks near the singularity or with time-development, although the radius grows in the solution of this paper. 
Therefore, the solution presented in this paper differs from the McVittie solution. 

We also consider the case where the EoS parameter is not constant. 
There are also solutions where the universe and the horizon oscillate. 

A new type of future singularity has also been discovered. 
We consider a spherical singularity. 
If the radius of the sphere vanishes, the singularity exists at the origin. 
If the horizon hides the singularity, cosmic censorship is not violated. 
If the radius becomes larger than the horizon radius, the naked singularity appears, which suggests that the cosmic censorship is violated. 
Note that the future singularities, like the Big Rip singularity, are naked singularities. 
If the radius of singularity becomes infinite, the universe could end as in the future singularity case, although the singularity is not inhomogeneous. 

In the next section, we consider spherically symmetric and time-dependent spacetime, especially the embedding of the black hole in the expanding universe. 
For later use, we calculate the explicit forms of the connections and curvatures. 
In Section~\ref{SecIII}, we consider the solution of the spherically symmetric and time-dependent spacetime for the Einstein equation by assuming the cosmic fluid with the corresponding EoS. 
We show that when the EoS parameter is a constant, the embedding of the Schwarzschild spacetime in the expanding universe is an exact solution of the Einstein equation. 
We consider the possibility that these black holes become candidates for the (super-)massive black hole and dark matter. 
After that, we consider a more general EoS and investigate several kinds of future singularities and an oscillating universe. 
The thermodynamics of the system where two heat sources appear as the black hole horizon and the cosmological horizon is studied. 
The effects of the black hole horizon on the tidal force generated by the expansion of the universe are discussed. 
The tidal force is enhanced by the horizon in some regions, but near the horizon, the tidal force in the radial direction acts to press the extended object, although 
the tidal forces in the angular directions act to tear the object. 
As a result, the extended object near the horizon might be crushed instead of being torn and ripped. 
In Section~\ref{SecIV}, we propose a new type of future singularity, where the black hole singularity becomes naked and extends over the universe. 
The structure of the spacetime is shown by using the Penrose diagram. 
The last section is devoted to the summary and discussions, including some remarks about the photon sphere. 


\section{Spherically symmetric and dynamical spacetime}\label{SecII}

Let us consider the spacetime which has spherical symmetry but is a dynamical one. 
Especially, we consider the embedding of the Schwarzschild solution in the background of the expanding universe. 
When the EoS parameter of the matter is a constant, we find exact solutions, which are different from the McVittie solution \cite{McVittie:1933zz}. 
We investigate the behaviour of the black hole and consider the role of the black hole in the expanding universe around the singularity. 
Especially, we discuss the tidal force, which acts on any extended object. 

\subsection{General Spherically symmetric and dynamical spacetime}\label{subsecIIA}

In this subsection, as a preparation, we consider the general time-dependent and spherically symmetric configuration of the spacetime, 
\begin{align}
\label{dsph}
ds^2 = - A(t,r) dt^2 + 2 B(t,r) dt dr + C(t,r) dr^2 + D(t,r) \left( d\theta^2 + \sin^2 \theta d\phi^2 \right)\, .
\end{align}
Such a spacetime can be realised by using two scalar fields \cite{Nojiri:2020blr}. 

In this paper, we use the following convention for the curvatures and connections, 
\begin{align}
\label{curvatures}
R=&\, g^{\mu\nu}R_{\mu\nu} \, , \quad R_{\mu\nu} = R^\lambda_{\ \mu\lambda\nu} \, , \quad 
R^\lambda_{\ \mu\rho\nu} = -\Gamma^\lambda_{\mu\rho,\nu} + \Gamma^\lambda_{\mu\nu,\rho} - \Gamma^\eta_{\mu\rho}\Gamma^\lambda_{\nu\eta} 
+ \Gamma^\eta_{\mu\nu}\Gamma^\lambda_{\rho\eta} \, ,\nonumber \\
\Gamma^\eta_{\mu\lambda} =&\, \frac{1}{2}g^{\eta\nu}\left( g_{\mu\nu,\lambda} + g_{\lambda\nu,\mu} - g_{\mu\lambda,\nu} \right)\, .
\end{align}
Because 
\begin{align}
\label{invg}
& g^{tt}= - \frac{C}{AC+B^2}\, , \quad g^{rr}= \frac{A}{AC+B^2}\, , 
\quad g^{tr}=g^{rt}= \frac{B}{AC+B^2}\, , \quad 
g^{\theta\theta} = \frac{1}{D} \, , \quad g^{\phi\phi}=\frac{1}{D \sin^2\theta} \, , \nonumber \\
& \mbox{other components of metric $g^{\mu\nu}$}=0 \, , 
\end{align}
we find 
\begin{align}
\label{Gmm}
\Gamma^t_{tt} =&\, \frac{C\dot A + 2B\dot B + BA'}{2(AC+B^2)} \, , \quad 
\Gamma^t_{tr} = \frac{CA' + B\dot C}{2(AC+B^2)} \, \quad 
\Gamma^r_{tt} = \frac{ - B\dot A + 2 A\dot B + AA'}{2(AC+B^2)}\, , \nonumber \\
\Gamma^r_{tr} =&\, \frac{-BA' + A\dot C}{2(AC+B^2)} \, , \quad 
\Gamma^t_{rr} = \frac{2CB' - C\dot C + BC'}{2(AC+B^2)} \, , \quad 
\Gamma^r_{rr} 
= \frac{2BB' - B \dot C + AC'}{2(AC+B^2)} \, , \nonumber \\
\Gamma^t_{\theta\theta} =&\, \frac{C\dot D - B D'}{2(AC+B^2)} \, , \quad 
\Gamma^t_{\phi\phi} 
= \frac{C\dot D - BD'}{2(AC+B^2)} \sin^2\theta \, , \quad 
\Gamma^\theta_{t\theta} = \Gamma^\theta_{\theta t} = \Gamma^\phi_{t\phi} = \Gamma^\phi_{\phi t} = \frac{\dot D}{2D} \, , \nonumber \\
\Gamma^r_{\theta\theta} =&\, - \frac{B\dot D + AD'}{2(AC + B^2)} \, , \quad 
\Gamma^r_{\phi\phi} = - \frac{B\dot D + AD'}{2(AC + B^2)} \sin^2\theta \, , \nonumber \\
\Gamma^\theta_{r\theta} =&\, \Gamma^\phi_{\phi r} = \frac{D'}{2D} \, , \quad 
\Gamma^\theta_{\phi\phi} = - \sin\theta \cos\theta \, , \quad 
\Gamma^\phi_{\theta\phi} = \Gamma^\phi_{\phi\theta} = \cot \theta \, , \quad \mbox{other components}=0\, .
\end{align}
Here $\dot A=\frac{\partial A}{\partial t}$ and $A' = \frac{\partial A}{\partial r}$, etc. 
The Ricci curvatures are given by, 
\begin{align}
\label{Ricci2}
R_{tt}=&\, - \frac{1}{2} \partial_t^2 \left( \ln \left( \left(AC+B^2\right) D^2 \right) \right) 
+ \partial_t \left\{ \frac{C\dot A + 2B\dot B + BA'}{2(AC+B^2)} \right\}
+ \partial_r \left\{ \frac{- B\dot A + 2 A\dot B + AA'}{2(AC+B^2)} \right\} \nonumber \\
&\, - \frac{(C\dot A + 2B\dot B + BA')^2}{4(AC+B^2)^2} - \frac{2(CA' + B\dot C)(- B\dot A + 2 A\dot B + AA')}{4(AC+B^2)^2} 
 - \frac{(-BA' + A\dot C)^2}{4(AC+B^2)^2} - \frac{{\dot D}^2}{2D^2} \nonumber \\
&\, + \frac{1}{2} \partial_t \left( \ln \left( \left(AC+B^2\right) D^2 \right) \right) \frac{C\dot A + 2B\dot B + BA'}{2(AC+B^2)} 
+ \frac{1}{2} \partial_r \left( \ln \left( \left(AC+B^2\right) D^2 \right) \right)\frac{- B\dot A + 2 A\dot B + AA'}{2(AC+B^2)} \, , \nonumber \\
R_{rr} =&\, - \frac{1}{2} \partial_r^2 \left( \ln \left( \left(AC+B^2\right) D^2 \right) \right) 
+ \partial_t \left\{ \frac{2CB' - C\dot C + BC'}{2(AC+B^2)} \right\} + \partial_r \left\{ \frac{2BB' - B \dot C + AC'}{2(AC+B^2)} \right\} \nonumber \\
&\, - \frac{(CA' + B\dot C)^2}{4(AC+B^2)^2} - \frac{2(-BA' + A\dot C)(2CB' - C\dot C + BC')}{4(AC+B^2)^2} 
 - \frac{(2BB' - B \dot C + AC')^2}{4(AC+B^2)^2} - \frac{{D'}^2}{2D^2} \nonumber \\
&\, + \frac{1}{2} \partial_t \left( \ln \left( \left(AC+B^2\right) D^2 \right) \right) \frac{2CB' - C\dot C + BC'}{2(AC+B^2)} 
+ \frac{1}{2} \partial_r \left( \ln \left( \left(AC+B^2\right) D^2 \right) \right) \frac{2BB' - B \dot C + AC'}{2(AC+B^2)} \, , \nonumber \\
R_{rt}=&\, R_{tr} \nonumber \\
=&\, - \frac{1}{2} \partial_t \partial_r \left( \ln \left( \left(AC+B^2\right) D^2 \right) \right) 
+ \partial_t \left\{ \frac{CA' + B\dot C}{2(AC+B^2)} \right\} + \partial_r \left\{ \frac{-BA' + A\dot C}{2(AC+B^2)} \right\} \nonumber \\
&\, - \frac{(CA' + B\dot C)(C\dot A + 2B\dot B + BA')}{4(AC+B^2)^2} - \frac{(-BA' + A\dot C)(CA' + B\dot C)}{4(AC+B^2)^2} \nonumber \\
&\, - \frac{(2BB' - B \dot C + AC')(-BA' + A\dot C)}{4(AC+B^2)^2} - \frac{\dot D D'}{2D^2} \nonumber \\
&\, + \frac{1}{2} \partial_t \left( \ln \left( \left(AC+B^2\right) D^2 \right) \right) \frac{CA' + B\dot C}{2(AC+B^2)}
+ \frac{1}{2} \partial_r \left( \ln \left( \left(AC+B^2\right) D^2 \right) \right) \frac{-BA' + A\dot C}{2(AC+B^2)} \, , \nonumber \\
R_{\theta\theta} =&\, 1 + \partial_t \left\{ \frac{C\dot D - B D'}{2(AC+B^2)} \right\} 
 - \partial_r \left\{ \frac{B\dot D + AD'}{2(AC + B^2)} \right\}
 - \frac{(C\dot D - B D')\dot D}{2(AC+B^2)D} + \frac{(B\dot D + AD') D'}{2(AC + B^2)D} \nonumber \\
&\, + \frac{1}{2} \partial_t \left( \ln \left( \left(AC+B^2\right) D^2 \right) \right) \frac{C\dot D - B D'}{2(AC+B^2)} 
 - \frac{1}{2} \partial_r \left( \ln \left( \left(AC+B^2\right) D^2 \right) \right) \frac{B\dot D + AD'}{2(AC + B^2)} \, , \nonumber \\
R_{\phi\phi} =&\, \partial_t \left\{ \frac{C\dot D - B D'}{2(AC+B^2)} \right\} \sin^2\theta 
 - \partial_r \left\{ \frac{B\dot D + AD'}{2(AC + B^2)} \right\} \sin^2\theta + \sin^2\theta \nonumber \\
&\, - \frac{(C\dot D - B D')\dot D}{2(AC+B^2) D} \sin^2\theta + \frac{(B\dot D + AD')D'}{2(AC + B^2)D} \sin^2\theta \nonumber \\
&\, + \frac{1}{2} \partial_t \left( \ln \left( \left(AC+B^2\right) D^2 \right) \right) \frac{C\dot D - B D'}{2(AC+B^2)} \sin^2\theta 
 - \frac{1}{2} \partial_r \left( \ln \left( \left(AC+B^2\right) D^2 \right) \right) \frac{B\dot D + AD'}{2(AC + B^2)} \sin^2\theta \, , \nonumber \\
&\, \mbox{other components of Ricci tensor} = 0 \, . 
\end{align}
Using the Einstein equation, 
\begin{align}
\label{Ee}
R_{\mu\nu} - \frac{1}{2} g_{\mu\nu} R = \kappa^2 T_{\mu\nu}\, , 
\end{align}
with a gravitational coupling $\kappa$, we find the energy-momentum tensor, that is, energy density, pressures, 
and their flows or currents, which realise the given geometry (\ref{dsph}). 

\subsection{Embedding of the black hole in the expanding universe}\label{subsecIIB}

In this subsection, as a time-dependent and spherically symmetric configuration of the spacetime, 
we consider the embedding of the black hole in the expanding universe. 

Let us start from the following metric, 
\begin{align}
\label{geo}
A=\e^{2N(t) + 2\nu(r)}\, , \quad B=0\, , \quad C=\e^{2N(t) + 2\lambda(r)}\, , \quad D=\e^{2N(t)} r^2\, ,
\end{align}
that is, the static and spherically symmetric spacetime embedded in the expanding universe, 
\begin{align}
\label{geo2}
ds^2 = \e^{2N(t)} \left\{ - \e^{2\nu(r)} dt^2 + \e^{2\lambda(r)} dr^2 + r^2 \left( d\theta^2 + \sin^2\theta d\phi^2 \right) \right\}\, ,
\end{align}
although the geometry might not always look natural but by this assumption, we can find exact solutions. 
 
The Hubble rate is defined by 
\begin{align}
\label{Hbbl}
H = \e^{-N(t)} \dot N\, ,
\end{align}
which is, in fact, the standard Hubble rate when $\nu=\lambda=0$. 
Then the connections are given by 
\begin{align}
\label{connections2}
& \Gamma^t_{tt} = \dot N \, , \quad 
\Gamma^r_{tt} = \e^{2\nu-2\lambda} \nu' \, , \quad 
\Gamma^t_{tr} = \Gamma^t_{rt} = \nu' \, , \quad 
\Gamma^t_{rr} = \e^{-2\nu + 2\lambda} \dot N \, , \quad 
\Gamma^r_{tr}=\Gamma^r_{rt} = \dot N \, , \nonumber \\
& \Gamma^t_{\theta\theta} = \dot N r^2 \, , \quad 
\Gamma^t_{\phi\phi} = \dot N r^2 \sin^2\theta \, , \quad 
\Gamma^\theta_{t\theta} = \Gamma^\theta_{\theta t} = \dot N\, , \quad 
\Gamma^\phi_{t\phi}=\Gamma^\phi_{\phi t} = \dot N\, , \quad 
\Gamma^r_{rr}=\lambda' \, , \nonumber \\
&\ \Gamma^r_{\theta\theta} = - r \e^{-2\lambda}\, , \quad 
\Gamma^r_{\phi\phi}= - r \e^{-2\lambda} \sin^2\theta \, , \quad 
\Gamma^\theta_{r\theta} = \Gamma^\theta_{\theta r} = r \, , \quad 
\Gamma^\phi_{r\phi} = \Gamma^\phi_{\phi r} = r \, , \quad 
\mbox{other components}=0 \, , 
\end{align}
and the Ricci curvature components are 
\begin{align}
\label{crvtrs2}
R_{tt}=&\, - 3 \ddot N + \e^{2\nu - 2\lambda} \left( \nu'' + {\nu'}^2 - \nu' \lambda' + \frac{2\nu'}{r} \right) \nonumber \\
R_{rr} =&\, \e^{-2\nu + 2\lambda}\left( \ddot N + 2 {\dot N}^2 \right) - \nu'' - {\nu'}^2 + \nu' \lambda' + \frac{2\lambda'}{r} + \frac{2}{r^2} \, , \nonumber \\
R_{rt}=&\, R_{tr} 
= - 2 \nu' \dot N - \lambda' \dot N - \frac{2 \dot N}{r} \, , \nonumber \\
R_{\theta\theta}=&\, r^2 \e^{-2\nu} \left( \ddot N + 2 {\dot N}^2 \right) + 1 - \left( \nu' - \lambda' + \frac{1}{r} \right) r \e^{-2\lambda} \, , \nonumber \\
R_{\phi\phi}=&\, r^2 \e^{-2\nu} \left( \ddot N + 2 {\dot N}^2 \right) \sin^2\theta + \sin^2\theta 
 - \left( \nu' - \lambda' + \frac{1}{r} \right) r \e^{-2\lambda} \sin^2\theta \, .
\end{align}
When $N=0$, we obtain the standard static and spherically symmetric spacetime. 
Furthermore, when we consider the spatially flat metric, that is, the case of $\nu=\lambda=0$, we obtain the standard 
Friedmann equations when we regard $t$ as a conformal time. 

The scalar curvature is given by 
\begin{align}
\label{scrvtr}
R=&\, \e^{-2N-2\nu} \left( 6\ddot N + 6{\dot N}^2 \right) + \e^{-2N-2\lambda} \left( -2 \nu'' - 2 {\nu'}^2 + 2\nu' \lambda' + \frac{- 4 \nu' + 4 \lambda'}{r} 
 - \frac{2}{r^2} \right) + \frac{2\e^{-2N}}{r^2} \, .
\end{align}
By using the energy-momentum tensor $T_{\mu\nu}$, we now define the energy density $\rho$, the pressure in the radial direction $p_\mathrm{rad}$, 
the pressure in the angular direction $p_\mathrm{ang}$, and energy flow $j$ by, 
\begin{align}
\label{rhopj}
T_{tt}=-\rho g_{tt}\, , \quad T_{rr} = p_\mathrm{rad} g_{rr}\, , \quad 
T_{\theta\theta} = p_\mathrm{ang} g_{\theta\theta} \, , \quad 
T_{\phi\phi}= p_\mathrm{ang} g_{\phi\phi} \, , \quad 
T_{tr}=T_{rt}= j \, .
\end{align}
Then the Einstein equation (\ref{Ee}) gives, 
\begin{align}
\label{Eeqs}
3{\dot N}^2 \e^{-2N-2\nu}+ \e^{-2N-2\lambda} \left( \frac{2\lambda'}{r} - \frac{1}{r^2} \right) + \frac{\e^{-2N}}{r^2} =&\, \kappa^2 \rho \, , \nonumber \\
\e^{-2N -2\nu} \left( - 2 \ddot N - {\dot N}^2 \right) + \e^{-2N - 2\lambda} \left( \frac{2\nu'}{r} + \frac{1}{r^2} \right) - \frac{\e^{-2N}}{r^2} =&\, \kappa^2 p_\mathrm{rad} \, , \nonumber \\
\e^{-2N-2\nu} \left( -2 \ddot N - {\dot N}^2 \right) - \e^{-2N - 2\lambda} \left( \nu'' + {\nu'}^2 - \nu'\lambda' + \frac{\nu' - \lambda'}{r} \right) =&\, \kappa^2 p_\mathrm{ang} 
\, , \nonumber \\
\dot N \left( - 2 \nu' - \lambda' - \frac{2}{r} \right) =&\, \kappa^2 j \, .
\end{align}
Therefore for arbitrary $N$, $\nu$, and $\lambda$, we find that $\rho$, $p_\mathrm{rad}$, and $p_\mathrm{ang}$ are given by the function of $t$ and $r$. 
This tells that if we delete $t$ and $r$ in the above expressions, we obtain a kind of equation of state (EoS) in the form of 
$f\left(\rho, p_\mathrm{rad}, p_\mathrm{ang}\right)=0$ in terms of an adequate function $f$. 

\section{General solution of spherically symmetric spacetime}\label{SecIII}

\subsection{Exact solution: Schwarzschild spacetime embedded in the expanding universe with a fluid with a constant equation of state parameter}\label{subsecIIIA}

In this subsection, we show that the Schwarzschild spacetime embedded in the universe, which is expanding due to a fluid with a constant 
EoS parameter $w$ as in (\ref{geo}), is an exact solution of the Einstein equation (\ref{Ee}). 

We now consider the case that $\nu$ and $\lambda$ are given by those of the Schwarzschild black hole, 
\begin{align}
\label{Schwrz}
\e^{2\nu}=\e^{-2\lambda}=1 - \frac{2M}{r}\, .
\end{align}
Here, $M$ is the mass of the black hole when $N=0$. 
Then the equations in (\ref{Eeqs}) become very simple, 
\begin{align}
\label{EeqsB}
3{\dot N}^2 \e^{-2N-2\nu} =&\, \kappa^2 \rho \, , \nonumber \\
\e^{-2N -2\nu} \left( - 2 \ddot N - {\dot N}^2 \right) =&\, \kappa^2 p_\mathrm{rad} \, , \nonumber \\
\e^{-2N-2\nu} \left( -2 \ddot N - {\dot N}^2 \right) =&\, \kappa^2 p_\mathrm{ang} 
\, , \nonumber \\
 - \dot N \left( \nu' + \frac{2}{r} \right) =&\, \kappa^2 j \, .
\end{align}
We should note $p_\mathrm{rad}=p_\mathrm{ang}$ as for the perfect fluid. 
If we define the effective energy density $\rho_\mathrm{eff}$ and the effective pressure $p_\mathrm{eff}$ by 
\begin{align}
\label{effrhp}
\rho_\mathrm{eff} \equiv \e^{2\nu} \rho\, , \quad p_\mathrm{eff} \equiv \e^{2\nu} p_\mathrm{rad} = \e^{2\nu} p_\mathrm{ang}\, ,
\end{align}
the first three equations in (\ref{EeqsB}) reduce to the standard first and second Friedmann equations by using the conformal time, 
\begin{align}
\label{EeqsBB}
3{\dot N}^2 \e^{-2N} =&\, \kappa^2 \rho_\mathrm{eff} \, , \nonumber \\
\e^{-2N} \left( - 2 \ddot N - {\dot N}^2 \right) =&\, \kappa^2 p_\mathrm{eff} \, . 
\end{align}
Therefore, if we assume the standard EoS of (\ref{EoS0}), as an example, given by 
\begin{align}
\label{EoS}
p_\mathrm{rad} = p_\mathrm{ang} = w \rho\, ,
\end{align}
with a constant EoS parameter $w$, we find
\begin{align}
\label{EoS2}
p_\mathrm{eff} = w \rho_\mathrm{eff}\, .
\end{align}
This indicates that if $-\frac{1}{3}>w>-1$, the black hole is embedded in the quintessence universe; 
if $w=-1$, it is embedded in the de Sitter universe; and if $w<-1$, it is embedded in the phantom universe. 
The energy flow $j$ in the last equation of (\ref{EeqsB}) expresses the accretion of matter generating the expansion of the black hole, accompanied by the expansion of the universe. 
The horizon exists at $r=2M$ as in the standard Schwarzschild metric. 
The areal radius is, however, given by 
\begin{align}
\label{ar}
\tilde r = \e^{N(t)} r \, ,
\end{align}
as clear from (\ref{geo2}). 
Then the radius of the horizon increases as $2M\e^{N(t)}$, that is, the radius is proportional to the expansion expressed by $\e^{N(t)}$. 

We should note that this solution is an exact solution of the Einstein equation, as in the McVittie metric \cite{McVittie:1933zz}, which is given by, 
\begin{align}
\label{McV}
ds^2 = - \left( \frac{1 - \frac{2m_0}{a\left( \tilde t \right)\tilde r}}{1 + \frac{2m_0}{a\left( \tilde t \right) \tilde r}} \right)^2 d{\tilde t}^2 
 - \left( 1 + \frac{2m_0}{a\left( \tilde t \right) \tilde r} \right) - \left( 1 + \frac{2 m_0}{a\left( \tilde t \right)\tilde r} \right)^4 a\left( \tilde t \right) 
\left\{ d{\tilde r}^2 + {\tilde r}^2 \left( d\theta^2 + \sin^2\theta d\phi^2 \right) \right\} \, .
\end{align}
Here $m_0$ is a constant, $\tilde t$, $\tilde r$, $a \left( \tilde t \right)$ are temporal and radial coordinates, and the scale factor in the McVittie metric. 
We should note that in the McVittie solution, the EoS parameter is not a constant but depends on the radial coordinate $\tilde r$. 
The horizon appears at $r= \frac{2m_0}{a \left( \tilde t \right)}$. 
If the EoS parameter $w$ is a constant, in the exact solution of this section, the black hole horizon grows by the accretion of the fluid or dark energy. 
On the other hand, in the McVittie solution (\ref{McV}), because when $a \left( \tilde t \right)$ is large as it becomes near the Type I or Big Rip singularity, the areal radius is given by 
$r_A\sim \sqrt{a\left( \tilde t \right)}\tilde r$, the horizon radius measured by the areal radius becomes smaller due to the expansion of the universe, 
$r_A \sim \frac{2m_0}{\sqrt{a\left( \tilde t \right)}}$. 
Therefore, the solution in this section is different from the McVittie solution, and it could be more natural because the EoS parameter is constant. 

In the spacetime far from the Schwarzschild black hole, $r\gg 2M$, in (\ref{Schwrz}), the universe is described by the spatially flat 
FLRW universe, where it is often convenient to use the cosmological time $\tau$ instead of the conformal time $t$. 
The relation between the proper time $\tau$ and the conformal time $t$ is given by 
\begin{align}
\label{rlcc} 
\e^{N(t)}dt = d\tau\, .
\end{align}
Because the Hubble rate defined by $H=\frac{dN}{d\tau}$ is rewritten by 
(\ref{Hbbl}), in the case of (\ref{EoS}) or (\ref{EoS2}), if $w>-1$, $H$ behaves as $H=\frac{\frac{2}{3(1+w)}}{\tau}$, if $w=-1$, $H$ is a constant corresponding to the de Sitter spacetime, 
and if $w<-1$, the behaviour of $H$ is given by the phantom universe $H=- \frac{\frac{2}{3(1+w)}}{\tau_\mathrm{BR} - \tau}$ with a constant $\tau_\mathrm{BR}$ corresponding to a 
Big Rip time. 
In this case, the Big Rip singularity occurs at $\tau=\tau_\mathrm{BR}$. 

Before proceeding, we clarify the behaviour of $N$ in terms of both $\tau$ and $t$. 
\begin{enumerate}
\item\label{i1} First, we consider the case $-\frac{1}{3}>w>-1$. 
As $H=\frac{dN}{d\tau}= \frac{\frac{2}{3(1+w)}}{\tau}$, we find $N=\frac{2}{3(1+w)} \ln \frac{\tau}{\tau_0}$. 
Here $\tau_0$ is a constant of the integration. 
By using (\ref{rlcc}), that is, $dt=\e^{-N(\tau)}d\tau$, we find 
$t = t_0 + \frac{3(1+w)}{1+3w} \left( \frac{\tau}{\tau_0} \right)^\frac{1+3w}{3(1+w)}$ and $N=N(t)= \frac{2}{1+3w} \ln \left( - \frac{1+3w}{3(1+w)}\left( t_0 - t \right) \right)$. 
Note $t<t_0$. 
The limit $t\to t_0$ corresponds to the Big Bang $\e^{N(t)}\to 0$. 
\item\label{i2} In the case of $w=-1$, as $H$ is a constant, $H=H_1$, $N=H_1 \left(\tau - \tau_0\right)$ with a constant $\tau_0$, 
$t=t_0 - \frac{1}{H_1} \e^{-H_1 \left(\tau - \tau_0\right)}$, and $N=N(t)= - \ln \left( H_1 \left( t_0 - t \right) \right)$. 
Here $t<t_0$. 
\item\label{i3} Finally when $w<-1$, we find $N=\frac{2}{3(1+w)} \ln \frac{\tau_\mathrm{BR} - \tau}{\tau_0}$ with a constant $\tau_0$, 
$t=t_0 - \frac{3(1+w)}{1+3w} \left( \frac{\tau_\mathrm{BR} - \tau}{\tau_0} \right)^\frac{1+3w}{3(1+w)}$, 
and $N=N(t)= \frac{2}{1+3w} \ln \left( \frac{1+3w}{3(1+w)}\left( t_0 - t \right) \right)$. 
We should note $t<t_0$, again, and the limit $\tau\to \tau_\mathrm{BR}$ corresponds to $t\to t_0$. 
\end{enumerate}
In the case \ref{i1}, that is, $-\frac{1}{3}>w>-1$, $t$ decreases if $\tau$ increases, that is, the direction of $t$ is opposite of $\tau$. 
On the other hand, in the cases \ref{i2} and \ref{i3}, that is, in the case of $w\leq -1$, we have chosen so that $t$ increases if $\tau$ increases. 

In the above exact solution of the Einstein equation, the mass and the radius of the horizon in the black hole increase associated with the expansion of the universe. 
This is due to the accretion of the fluid, which could be dark energy. 
The accretion is described by $j$ in (\ref{EeqsB}), which is the flow of energy in the radial direction. 
When $-\frac{1}{3}>w>-1$, we find 
\begin{align}
\label{wl}
j= \frac{2\left(2r - 3M\right)}{\kappa^2 \left(1+3w\right)\left( t_0 - t\right) r \left( r - 2M \right)}\, .
\end{align}
Outside the horizon $r>2M$, $j$ is negative because $1+3w<0$ and $t<t_0$, which tells that the fluid falls into the black hole. 
In the case $w=-1$, we obtain, 
\begin{align}
\label{we}
j = - \frac{2 r - 3M}{\kappa^2 \left(t_0 - t\right)r\left( r - 2M\right)} \, .
\end{align}
When $t<t_0$, $j$ is negative outside the horizon again. 
For $w<-1$ case, $j$ is given by 
\begin{align}
\label{wg}
j = \frac{2\left(2r - 3M\right)}{\kappa^2 \left(1+3w\right)\left( t_0 - t\right) r \left( r - 2M \right)}\, .
\end{align}
Again, $j$ is negative outside the horizon. 
We should note that $j$ diverges at the Big Rip time $t=t_0$, where the radius of the horizon also becomes infinite. 

{\color{red}
Although we are interested in the behaviour of the asymptotically acceleratedly expanding universe, we may also consider a critical value of the EoS parameter $w=-\frac{1}{3}$ 
because the expression in (\ref{wl}) diverges in the limit of $w\to - \frac{1}{3}$. 
The critical value corresponds to the boundary value of $w$ between the accelerating expansion and the decelerating expansion of the universe. 
When $w=-\frac{1}{3}$, because $H=\frac{dN}{d\tau}= \frac{1}{\tau}$, we find $N=\ln\frac{\tau}{\tau_0}$ with a constant of the integration $\tau_0$. 
By using the relation $dt=\e^{-N(\tau)}d\tau$, we obtain $t=t_0 + \ln \frac{\tau}{\tau_0}$ with a new constant of the integration $t_0$ and we also find 
$N(t)=t-t_0$. 
Then the last equation of Eq.~(\ref{EeqsB}) tells 
\begin{align}
\label{j13rd}
j = - \frac{2r - 3M}{\kappa^2 r\left(r-2M\right)} \, .
\end{align}
Therefore, for the critical value $w=-\frac{1}{3}$, the flow $j$ of the energy in the radial direction does not depend on time. 
Eq.~(\ref{j13rd}) can be also obtain by choosing $t_0=\frac{2}{1+3w}$ in (\ref{wl}) and taking the limit of $w\to -\frac{1}{3}$. 
}

Just before the Big Rip singularity, the tidal force pulls and eventually tears any bounded objects apart. 
Because the horizon is not an object but the structure of spacetime, its size is merely enlarged by the expansion and becomes infinite at the singularity, as shown in this section. 
The enlargement of the horizon is also understood in terms of the flow of energy in (\ref{wg}), which diverges at the singularity because the energy density of the fluid, 
which is phantom, becomes infinite at the singularity. 

\subsection{Possible candidate of black holes in GW231123}

Ref.~\cite{LIGOScientific:2025rsn} reported on the gravitational wave event GW231123, which is consistent with the merger of two black holes with masses 
$137^{+22}_{-17}\, M_\odot$ and $103^{+20}_{-52}\, M_\odot$. 
Here $M_\odot$ is the solar mass. 
It is believed that the black holes with these mass scales could not be created by the collapse of astrophysical objects.
In the solution of this paper (\ref{geo}) with (\ref{Schwrz}), the black hole grows due to the expansion of the universe. 
Therefore, if there are primordial black holes \cite{Carr:1975qj, Carr:2020gox, Sasaki:2018dmp, Papanikolaou:2025ddc} during inflation, these black holes become larger and larger and 
large black holes could remain after inflation. 
This scenario may also explain the origin of the black holes in GW231123 and supermassive black holes at the centre of galaxies. 
Moreover, these black holes may be candidates for dark matter.

In our model (\ref{geo}) with (\ref{Schwrz}), the horizon exists at $r=r_\mathrm{h}=2M$ for the radial coordinate $r$. 
On the other hand, the areal radius is given by (\ref{ar}), and we find that the horizon radius ${\tilde r}_\mathrm{h}$ for the areal coordinate is 
\begin{align}
\label{GW}
{\tilde r}_\mathrm{h} = \e^{N(t)} r_\mathrm{h} = 2 \e^{N(t)} M\, .
\end{align}
The black hole mass $M_\mathrm{BH}$ grows exponentially due to the expansion of the universe, 
\begin{align}
\label{GW2}
M_\mathrm{BH} = \e^{N(t)} M\, .
\end{align}
If the $e$-folding number of the inflation is $N=60$, because $\e^{60}\sim 10^{26}$, the masses of the primordial black holes created before the inflation increase by 
$10^{26}$ times, and the masses of the primordial black holes created during inflation also increase significantly. 
Microscopic primordial black holes are believed not to live for a long time, and they could evaporate due to the 
Hawking radiation, but due to the above mechanism, they might survive in the present universe, and they might be a part of dark matter. 
This mechanism also produces massive black holes, as suggested by the GW231123 event and supermassive black holes at the centre of galaxies. 

The solar mass is given by $M_\odot \approx 1.9885 \times 10^{30} \, \mathrm{kg}$. 
If the light primordial black holes, which might be dark matter with the typical mass $10^{-16}-10^{-10} \, M_\odot$, exist before the inflation, 
the masses in that time could be $10^{-42}-10^{-36} \, M_\odot \sim 10^{-12}-10^{-6}\, \mathrm{kg}$, which are small but much larger than the mass of elementary particles 
like proton, whose mass is about $10^{-27}\, \mathrm{kg}$. 
As the Hawking temperature of the black hole with the mass identical to the sun is $T_\mathrm{H} \approx 6.17 \times 10^{-8} \, \mathrm{K}$, 
the Hawking temperatures of the primordial black hole before the inflation could be $T_\mathrm{H} \approx 10^{34}-10^{28} \, \mathrm{K}$, which could be very high. 
Therefore, if the period between the creation of the primordial black hole and the inflation is not very short or the primordial black holes are not created during the inflation, 
they may not survive in the present universe due to the Hawking evaporation. 

On the other hand, if the massive black holes with $10^2\, M_\odot$ suggested by the GW231123 event were created before the inflation, the masses could be 
$10^{-24}\, M_\odot \sim 10^6\, \mathrm{kg}$. 
Therefore, these black holes might be created before inflation due to the fluctuation, although it depends on the corresponding scenario or models of the primordial black hole creation. 
In the case of supermassive black holes at the centre of the galaxies, whose typical mass ranges $10^6 - 10^{10}\, M_\odot$, the mass before the inflation could be 
$10^{-20} - 10^{-16}\, M_\odot$, which could not be so large. 
These black holes may be created by the fluctuations before the inflation. 

\subsection{More general EoS and future singularity}\label{subsecIIIB}

We now consider more general EoS than (\ref{EoS}) or (\ref{EoS2}) and the models with different future singularities. 

As clear from (\ref{EeqsBB}), $\rho_\mathrm{eff}$ and $p_\mathrm{eff}$ satisfy the conservation law, 
\begin{align}
\label{cnslw}
0= \frac{d \rho_\mathrm{eff}}{d\tau} + 3H \left( \rho_\mathrm{eff} + p_\mathrm{eff} \right)\, .
\end{align}
Therefore if $\rho_\mathrm{eff}$ and $p_\mathrm{eff}$ satisfy the following EoS, 
\begin{align}
\label{EoS3}
p_\mathrm{eff} = - \rho_\mathrm{eff} + f\left( \rho_\mathrm{eff} \right)\, ,
\end{align}
with a function $f\left( \rho_\mathrm{eff} \right)$, the conservation law (\ref{cnslw}) can be solved as 
\begin{align}
\label{Nrho}
N = - \int \frac{d\rho_\mathrm{eff}}{3f\left( \rho_\mathrm{eff} \right)}\, ,
\end{align}
and by using (\ref{Hbbl}), the first equation in (\ref{EeqsBB}) can be rewritten as 
\begin{align}
\label{1F}
\frac{1}{\kappa \sqrt{3\rho_\mathrm{eff}}f\left( \rho_\mathrm{eff} \right)}\frac{d\rho_\mathrm{eff}}{d\tau} = 1\, .
\end{align}
The second equation in (\ref{EeqsBB}) can be obtained from the first equation in (\ref{EeqsBB}) and the conservation law (\ref{cnslw}). 
By solving (\ref{1F}), we obtain the $\tau$ dependence of $\rho_\mathrm{eff}$. 
By using the obtained expression of $\rho_\mathrm{eff}=\rho_\mathrm{eff}(\tau)$, we obtain $N=N(\tau)$ by using (\ref{Nrho}). 

On the other hand, if we know the $\tau$ dependence of $H(\tau)$, we may solve the expression with respect to $H$ as $\tau=\tau(H)$. 
Then the equations (\ref{EeqsBB}) give 
\begin{align}
\label{f}
f\left( \rho_\mathrm{eff} \right) = - \frac{2}{\kappa^2} \left. \frac{dH}{d\tau} \right|_{\tau=\tau \left( H=\kappa \sqrt{\frac{\rho_\mathrm{eff}}{3}} \right)}\, .
\end{align}
Therefore, one can find $f\left( \rho_\mathrm{eff} \right)$ corresponding to an arbitrary $H(\tau)$. 

Now we write $p_\mathrm{rad} = p_\mathrm{ang} = p$ as in the perfect fluid. 
Then by using (\ref{effrhp}), Eq.~(\ref{EoS3}) shows 
\begin{align}
\label{EoS4}
p = - \rho + \e^{-2\nu(r)} f\left( \rho \e^{-2\nu(r)} \right)\, .
\end{align}
That is, the EoS becomes $r$-dependent or $\nu$ (or $\lambda$ because $\lambda=-\nu$)-dependent, as in the McVittie solution, 
except in the case of (\ref{EoS}). 

Because the spacetime far from the Schwarzschild black hole in (\ref{Schwrz}) is the spatially flat FLRW universe, we review the future singularity appearance in the FLRW universe. 
For the spatially flat FLRW universe, whose metric is given by 
\begin{align}
\label{FLRW}
ds^2 = - d\tau^2 + a(\tau)^2 \sum_{i=1,2,3} \left( dx^i \right)^2 \, ,
\end{align}
with a scale factor $a(\tau)$, the types of future singularities in cosmology are classified as follows \cite{Nojiri:2005sx}, when $t\rightarrow t_0$ ($\tau\to \tau_0$ or $\tau_\mathrm{BR}$), 
\begin{itemize}
\item Type I (Big Rip) singularity: 
$a(\tau)\rightarrow \infty$, $\rho\rightarrow\infty$ and $|p|\rightarrow\infty$.
\item Type II (sudden) singularity: 
$a(\tau) \rightarrow \mathrm{const.}$ and $\rho \rightarrow \mathrm{const.}$, but $|p|\rightarrow\infty$.
$a(\tau)$ and $\dot a(\tau)$ are finite, but $\ddot a(\tau)$ diverges. 
\item Type III (big freeze) singularity: 
$a(\tau) \rightarrow \mathrm{const.}$, but $\rho\rightarrow\infty$ and $|p|\rightarrow\infty$. 
$a(\tau)$ is finite, but $\dot a(\tau)$ diverges.
\item Type IV (generalised sudden) singularity: 
$a(\tau) \rightarrow \mathrm{const.}$, $\rho \rightarrow \mathrm{const.}$, and $|p| \rightarrow \mathrm{const.}$, but some higher derivatives of $H$ diverge.
$a(\tau)$, $\dot a(\tau)$, and $\ddot a(\tau)$ are finite, but higher derivatives of $a(\tau)$ diverge.
\end{itemize}
In addition to the above four types of singularities, the following has been proposed \cite{Dabrowski:2009kg}, 
\begin{itemize}
\item Type V ($w$) singularity: 
$w \rightarrow \infty$, but $p$ and $\rho$ are finite.
\end{itemize}
In Type V singularity, the behaviour of $a(\tau)$ is identical to that in Type II, 
that is, $a(\tau)$ and $\dot a(\tau)$ are finite, but $\ddot a(\tau)$ diverges. 
Type V singularity is just for the behaviour of the matter. 
Type I-IV singularities have completely classified the singular behaviours of spacetime. 

Type I singularity was first introduced in~\cite{Caldwell:2003vq}, which appears in the universe filled by phantom fluid~\cite{Caldwell:1999ew}. 
Type II singularity was proposed in~\cite{Barrow:2004xh}. 
Type III and Type IV singularities were obtained by complementing the Type I and Type II singularities in~\cite{Nojiri:2005sx}. 

Let us assume that a singularity appears at $\tau=\tau_s$ ($\tau_s=\tau_0$ or $\tau_s=\tau_\mathrm{BR}$). 
When $H$ behaves as $H\sim H_1 \left( \tau_s - \tau \right)^\alpha$ with constants $H_1>0$ and $\alpha$, 
if $\alpha\leq -1$, Type I singularity appears and if $-1<\alpha<0$, Type III appears. 
When $H\sim H_1 + H_2 \left( \tau_s - \tau \right)^\alpha$ with constants $H_1>0$, $H_2$, and $\alpha$, if $0<\alpha<1$, Type II singularity appears, 
and if $\alpha>1$ but $\alpha$ is not an integer, Type IV singularity apears. 

In the case of Type I and III singularities, the geodesics of the particles end at the singularity, and therefore, the singularities correspond to the end of the universe. 
On the other hand, in the cases of Type II and IV singularities, the geodesic could go through the singularities, and we may consider the universe after the singularities as 
in \cite{Odintsov:2022umu}, where the relation of the singularities with the Hubble tension or 
$H_0$ tension \cite{DiValentino:2021izs, Schoneberg:2021qvd} and the black hole shadow after the singularities have been discussed. 
The universe could exist after the Type II and IV singularities if the Hubble rate is given by $H\sim H_1 + H_2 \left| \tau_s - \tau \right|^\alpha$ in terms of 
the absolute value $\left| \tau_s - \tau \right|^\alpha$ instead of $H\sim H_1 + H_2 \left( \tau_s - \tau \right)^\alpha$ because $\left( \tau_s - \tau \right)^\alpha$ has 
a branch cut at $\tau=\tau_s$, which makes $\left( \tau_s - \tau \right)^\alpha$ ill-defined when $\tau>\tau_s$. 
Instead of using the absolute value, when $\alpha$ is given by $\alpha=\frac{k}{2l+1}$ by using two integers, we may choose the branch cut  
$\left( \tau_s - \tau \right)^\alpha$ as follows, 
\begin{align}
\label{alpha}
\left( \tau_s - \tau \right)^{\alpha =\frac{k}{2l+1}} = \left\{ 
\begin{array}{cc} 
\left( \tau_s - \tau \right)^{\alpha =\frac{k}{2l+1}} & \mbox{when}\ \tau<\tau_s \\
0 & \mbox{when}\ \tau=\tau_s \\
\left\{ - \left( \tau - \tau_s\right)^\frac{1}{2l+1} \right\}^k & \mbox{when}\ \tau>\tau_s
\end{array}
\right. \, .
\end{align}
By using the definition (\ref{alpha}), we may consider the universe when $\tau>\tau_s$. 

When $H\sim H_1 \left( \tau_s - \tau \right)^\alpha$, we have $\frac{dH}{d\tau}\sim - H_1 \alpha \left( \tau_s - \tau \right)^{\alpha-1}$ and 
$\tau \sim \tau_s - \left( \frac{H}{H_1} \right)^\frac{1}{\alpha} = \tau_s - \left( \frac{\kappa}{H_1} \sqrt{\frac{\rho_\mathrm{eff}}{3}} \right)^\frac{1}{\alpha}$. 
Therefore by using (\ref{f}), we find
\begin{align}
\label{I_III}
f\left( \rho_\mathrm{eff} \right) = \frac{2H_1\alpha}{\kappa^2} \left( \frac{\kappa}{H_1} \sqrt{\frac{\rho_\mathrm{eff}}{3}} \right)^\frac{\alpha - 1}{\alpha} \, .
\end{align}
The EoS for $\rho$ and $p=p_\mathrm{rad} = p_\mathrm{ang}$ is given by using (\ref{EoS4}). 
Then in the spacetime embedded the Schwarzschild black hole, if $\alpha\leq -1$, Type I singularity appears and if $-1<\alpha<0$, Type III appears. 

On the other hand, when $H\sim H_1 + H_2 \left( \tau_s - \tau \right)^\alpha$, $\frac{dH}{d\tau}\sim - H_2 \alpha \left( \tau_s - \tau \right)^{\alpha-1}$ and 
$\tau \sim \tau_s - \left( \frac{H-H_1}{H_2} \right)^\frac{1}{\alpha} = \tau_s - \left\{ \frac{1}{H_2} \left(\kappa\sqrt{\frac{\rho_\mathrm{eff}}{3}} - H_1 \right) \right\}^\frac{1}{\alpha}$. 
we obtain
\begin{align}
\label{II_IV}
f\left( \rho_\mathrm{eff} \right) = \frac{2H_2\alpha}{\kappa^2} \left\{ \frac{1}{H_2} \left(\kappa\sqrt{\frac{\rho_\mathrm{eff}}{3}} - H_1 \right) \right\}^\frac{\alpha - 1}{\alpha}\, .
\end{align}
Then if $0<\alpha<1$, Type II singularity appears and if $\alpha>1$ but $\alpha$ is not an integer, Type IV singularity appears. 
The EoS for $\rho$ and $p=p_\mathrm{rad} = p_\mathrm{ang}$ is given by using (\ref{EoS4}), again. 

Eq.~(\ref{rlcc}) and the last expression of (\ref{EeqsB}) show
\begin{align}
\label{EeqsC}
 - \e^{N(\tau)} \frac{dN}{d\tau} \left( \nu' + \frac{2}{r} \right) =&\, \kappa^2 j \, .
\end{align}
In the case of Type I singularity, because $a(\tau)=\e^{N(\tau}) \to \infty$ and $\frac{dN}{d\tau} \to \infty$ at the singularity, 
$j$ diverges, which is consistent with the result in the last section for the Big Rip singularity corresponding to $w<-1$. 
In the case of Type II singularity, $\dot a(\tau)= \e^{N(\tau)} \frac{dN}{d\tau}$ is finite and therefore $j$ is finite but $\frac{dj}{d\tau}$ diverges at the singularity. 
In the case of Type III singularity, although $a(\tau)$ is finite but $\frac{dN}{d\tau} \to \infty$ at the singularity, $j$ diverges. 
In the case of Type IV singularity, $j$ is finite, but the higher derivative of $j$ diverges. 

As in \cite{Nojiri:2006ww}, we may consider the case that the background universe is oscillating. 
In the paper \cite{Nojiri:2006ww}, a simple example is given as in \cite{Dodelson:2000jtt}, 
\begin{align}
\label{T8}
w=\frac{p_\mathrm{eff}}{\rho_\mathrm{eff}} = -1 + w_0 \cos \omega \tau\, .
\end{align}
Here, $w_0$ and $\omega$ are positive constants. 
Then we find, 
\begin{align}
\label{T9}
H = \frac{2\omega}{3\left(w_1 + w_0 \sin \omega \tau\right)}\, .
\end{align}
Here, $w_1$ appears as a constant of the integration. 
If $|w_1|<w_0$, the denominator can vanish, and the Type I or Big Rip singularity occurs. 
On the other hand, if $|w_1|>w_0$, such a singularity does not appear. 
By following \cite{Nojiri:2006ww}, by eliminating the time coordinate $\tau$, we obtain the equation of state for the ideal fluid, 
\begin{align}
\label{T13}
{w_0}^2 = \left( w_1 - \frac{2\omega}{\kappa\sqrt{3\rho_\mathrm{eff}}} \right)^2
+ \left(1 + \frac{p_\mathrm{eff}}{\rho_\mathrm{eff}}\right)^2 \, .
\end{align}
In this model, the oscillation of the background universe generates the oscillation of the horizon radius of the black hole. 


\subsection{Thermodynamics}

Let us consider thermodynamics. 
By using the surface gravity $\mathcal{K}$, the Hawking temperature $T_\mathrm{H}$ is given by 
\begin{align}
\label{sg}
T_\mathrm{H} = \frac{\mathcal{K}}{2\pi}\, .
\end{align}
When the metric has the following form, 
\begin{align}
\label{rs}
ds^2 = - f_\mathrm{red\, shift} (r)\, d\tau^2 + \cdots\, ,
\end{align}
with the redshift factor $f_\mathrm{red\, shift} (r)$, the surface gravity $\mathcal{K}$ is given by, 
\begin{align}
\label{sg2}
\mathcal{K}=\frac{1}{2} \frac{df_\mathrm{red\, shift} \left( r_\mathrm{h} \right)}{d\tilde r}\, .
\end{align}
Here $r_\mathrm{h}=2M$ is the value of the radial coordinate at the horizon, and $\tilde r$ is the areal radius in (\ref{ar}) for the metric choice (\ref{geo2}). 
By using the proper time $\tau$ defined by (\ref{rlcc}) and the embedding of the Schwarzschild black hole (\ref{Schwrz}), 
$f_\mathrm{red\, shift} (r) = \e^{2\nu(r)} = 1 - \frac{2M}{r}$, we find 
\begin{align}
\label{sg3}
\mathcal{K}=\frac{\e^{-N(\tau)}}{4M}\, , 
\end{align}
and
\begin{align}
\label{sg4}
T_\mathrm{H}= \frac{\e^{-N(\tau)}}{8\pi M}\, . 
\end{align}
Note that the Hawking temperature decreases due to the expansion of the universe, which is natural because the black hole grows due to the expansion in our solution. 
If we consider the Bekenstein-Hawking entropy \cite{Bekenstein:1972tm, Hawking:1974rv}, the black hole entropy $\mathcal{S}$ is given by 
\begin{align}
\label{ent}
\mathcal{S}=\frac{4\pi \e^{2N(\tau)} M^2}{G}=\frac{1}{16\pi {T_\mathrm{H}}^2}\, .
\end{align}
The last expression is a standard one. 
The entropy increases due to the expansion of the universe. 

On the other hand, in the expanding universe, there appears a cosmological horizon, which is an apparent horizon, whose radius is given by 
\begin{align}
\label{chr}
r_\mathrm{h\, cosmo}=\frac{1}{H}\, .
\end{align}
This tells that the entropy $\mathcal{S}_\mathrm{cosmo}$ given by the cosmological horizon has the following form, 
\begin{align}
\label{CosEnt}
\mathcal{S}_\mathrm{cosmo} = \frac{\pi}{G H^2}\, .
\end{align}
and the Hawking temperature of the cosmological horizon has the following form, 
\begin{align}
\label{chT}
T_\mathrm{H\, cosmo} = \frac{H}{2\pi}=\frac{1}{2\pi} \frac{dN}{d\tau}\, .
\end{align}
We have two heat sources, that is, the black hole and the cosmological horizon, which show that there are flows of heat and fluid between the two sources. 

We should note that Eq.~(\ref{ent}) shows that the black hole entropy monotonically increases with the expansion of the universe. 
That occurs because the black hole absorbs the surrounding fluid. 
On the other hand, the cosmological entropy (\ref{CosEnt}) decreases if the Hubble rate $H$ increases as in the Type I, II, and III singularities. 
This could indicate that there are flows of heat and fluid from the cosmological horizon to the black hole horizon. 

In general, the temperature of the black hole $T_\mathrm{H}$ is different from the temperature of the cosmological horizon 
$T_\mathrm{H\, cosmo}$ and the system is in the thermodynamical non-equilibrium. 
If $T_\mathrm{H} = T_\mathrm{H\, cosmo}$, that is, 
\begin{align}
\label{phstr}
\frac{\e^{-N(\tau)}}{8\pi M}=\frac{1}{2\pi} \frac{dN}{d\tau} \, ,
\end{align}
the phase transition will occur as in the Hawking-Page phase transition \cite{Hawking:1982dh} at the non-zero temperature. 
This is the stability on the semi-classical or quantum corrected level. 
When $T_\mathrm{H} > T_\mathrm{H\, cosmo}$, the black hole will evaporate but when $T_\mathrm{H} < T_\mathrm{H\, cosmo}$, due to the heat flux from 
the cosmological horizon, the black hole does not evaporate, but it will become thermodynamically stable. 
When the Hubble rate $H$ has a finite lower bound, this always occurs in the expanding universe due to the factor $\e^{-N(\tau)}$ in (\ref{sg4}), 
which becomes unboundedly smaller by the expansion of the universe. 

One more interesting point is that if 
\begin{align}
\label{rds}
2M \e^{2N} > \frac{1}{H}\, ,
\end{align}
the radius of the cosmological horizon $r_\mathrm{h\, cosmo}$ is smaller than the radius of the black hole horizon ${\tilde r}_\mathrm{h}=\e^{N(t)} r_\mathrm{h}$. 
If the diameter of the black hole horizon is larger than the radius of the cosmological horizon, 
\begin{align}
\label{rds2}
4M \e^{2N} > \frac{1}{H}\, ,
\end{align}
If the distance between the two points near the black hole horizon is larger than the radius of the cosmological horizon, the two points are causally disconnected from each other 
even if the points are outside the black hole horizon. 
More exactly, we need to reconsider the causal structure when $M \e^{2N} \sim \frac{1}{H}$. 
It is not so clear what could happen when two null surfaces look to merge, as in this case. 

We now consider what could happen when the future singularities occur. 
\begin{itemize}
\item For Type I singularity, before the singularity, although the entropy of the cosmological horizon becomes rapidly small, the black hole entropy grows. 
Before the Type I singularity, the black hole becomes stable because the Hawking temperature of the black hole becomes smaller than the Hawking temperature of 
the cosmological horizon. 
The radius of the cosmological horizon becomes much smaller than the black hole horizon. 
\item For Type II singularity, $N$ and the Hubble rate are finite at the singularity. 
Although the black hole temperature becomes smaller than the cosmological temperature and 
the black hole entropy becomes larger than the cosmological entropy, finally, such transitions may occur before the Type II singularity or 
after the singularity, depending on the value of $M$. 
The radius of the cosmological horizon also becomes smaller than the radius of the black hole horizon, finally, but the time of the transition also depends on $M$. 
\item For Type III singularity, because $N$ is finite but $H$ diverges at the singularity, the temperature $T_\mathrm{H}$ (\ref{sg4}), the entropy 
$\mathcal{S}_\mathrm{H}$ (\ref{ent}), and the horizon radius ${\tilde r}_\mathrm{h}=\e^{N(t)} r_\mathrm{h}$ of the black hole are finite but 
the temperature $T_\mathrm{H\, cosmo}$ (\ref{chT}), the entropy $\mathcal{S}_\mathrm{cosmo}$ (\ref{CosEnt}), and the radius $r_\mathrm{h\, cosmo}$ (\ref{chr}) 
of the horizon go to vanish. 
Therefore, as for Type I singularity, before Type III singularity, the entropy of the cosmological horizon becomes rapidly much smaller than that of the black hole entropy, 
the black hole becomes stable, and the radius of the cosmological horizon becomes much smaller than the black hole horizon. 
\item For Type IV singularity, the qualitative behaviours of the temperature, entropy, and the horizon radius are not changed from those in Type II singularity. 
Depending on the value of $M$, the black hole temperature becomes smaller than the cosmological temperature, 
the black hole entropy becomes larger than the cosmological entropy, and the radius of the cosmological horizon becomes smaller than 
the radius of the black hole horizon, finally, but such transitions may occur before the Type IV singularity or after the singularity. 
\end{itemize}


\subsection{Extended object and particle near the singularity}\label{subsecIIIC}

In this subsection, we investigate what could happen near the singularities following Ref.~\cite{Perivolaropoulos:2016nhp}. 
Especially, we find how the horizon could affect the behaviour of the extended or finite-size objects. 

First, we consider the geodesic deviation equation \cite{Wald:1984rg}, which describes how nearby geodesics (i.e., paths of free-falling particles) 
diverge or converge due to spacetime curvature, 
\begin{align}
\label{geodesicdev}
\frac{D^2 S^\mu}{d{\tilde\tau}^2} = R^\mu_{\ \nu\rho\sigma}T^\nu T^\rho S^\sigma \, .
\end{align}
Here, ${\tilde\tau}$, $S^\mu$, and $T^{\mu}$ are the proper time, deviation vector, and the tangent vector, respectively. 

In the case of the FLRW universe in (\ref{FLRW}), by choosing $T^\tau=1$, $T^i=0$, 
Eq.~(\ref{geodesicdev}) is reduced into the following form, 
\begin{align}
\label{geodesicdev1}
\frac{D^2 S^i}{d\tau^2} = R^i_{\ \tau\tau j} S^j \, .
\end{align}
Due to 
\begin{align}
\label{Riemann}
R^{i}_{\ \tau\tau j} = \left( \frac{dH}{d\tau} + H^2 \right) \delta_{ij} \, ,
\end{align}
Eq.~(\ref{geodesicdev1}) gives 
\begin{align}
\label{geodesicdev2}
\frac{D^2 S^i}{d\tau^2} = \left( \frac{dH}{d\tau} + H^2 \right) S^i \, .
\end{align}
$H$ and $\frac{dH}{d\tau}$ diverge in Type I and III singularities, and $\frac{dH}{d\tau}$ diverges in Type II singularity. 
Thus, Eq.~(\ref{geodesicdev2}) shows that spacetime is torn and ripped. 
If $H$, $\frac{dH}{d\tau}$, or both go to infinity in the infinite future, everything is ripped finally, which is a Little Rip~\cite{Frampton:2011sp, Brevik:2011mm, Frampton:2011rh}. 

For the spacetime described by (\ref{geo2}), we choose $T^t=\e^{-N - \nu}$, $T^i=0$ $\left(i=r,\theta,\phi\right)$. 
We first consider the angular directions, $S^\theta$ and $S^\phi$. 
Using 
\begin{align}
\label{Rthttht}
R^\theta_{\ t\theta t} = R^\theta_{\ t\theta t} = - \ddot N \, ,
\end{align}
we obtain, 
\begin{align}
\label{geodesicdev1thph}
\frac{D^2 S^\theta}{d{\tilde\tau}^2} = R^\theta_{\ tt\theta} \e^{-2N - 2\nu} S^r 
= - R^\theta_{\ t\theta t} \e^{-2N - 2\nu} S^\theta 
= \frac{\ddot N \e^{-2N}}{1 - \frac{2M}{r}} S^\theta \, , \quad 
\frac{D^2 S^\phi}{d{\tilde\tau}^2} 
= \frac{\ddot N \e^{-2N}}{1 - \frac{2M}{r}} S^\phi \, .
\end{align}
 From (\ref{Hbbl}), one finds 
\begin{align}
\label{lr2}
\dot N = \frac{dN}{dt} = H \e^N\, , \quad 
\ddot N = \left( \frac{dH}{d\tau} + H^2 \right) \e^{2N}\, , 
\end{align}
By using $H$, we rewrite (\ref{geodesicdev1thph}) as follows, 
\begin{align}
\label{geodesicdev1thph2}
\frac{D^2 S^{\theta,\phi}}{d{\tilde\tau}^2} 
= \frac{\frac{dH}{d\tau} + H^2}{1 - \frac{2M}{r}} S^{\theta,\phi} \, .
\end{align}
By comparing the above expression with (\ref{geodesicdev2}), we find that in Type I, II, and III singularities, any extended object is torn and ripped by the tidal force. 
The tidal force is enhanced near the horizon $r\to 2M$ due to the factor $\left( 1 - \frac{2M}{r} \right)^{-1}$. 

We also consider the deviation in $r$-direction, $S^r$. 
Then, Eq.~(\ref{geodesicdev}) has the following form, 
\begin{align}
\label{geodesicdev1B}
\frac{D^2 S^r}{d{\tilde\tau}^2} = R^r_{\ ttr} \e^{-2N - 2\nu} S^r 
= R^t_{\ rtr} \e^{-2N -2\lambda} S^r \, .
\end{align}
For the spacetime (\ref{geo2}), the Riemann tensor $R^t_{\ rtr}$ is given by 
\begin{align}
\label{Rietrtr2}
R^t_{\ rtr} = - \nu'' - 2{\nu'}^2 + \e^{-2\nu+2\lambda} \ddot N + \e^{-2\nu+2\lambda} \dot N \lambda' \, .
\end{align}
Further by using (\ref{Schwrz}), we find
\begin{align}
\label{Rietrtr3}
R^t_{\ rtr} = \frac{\frac{2M}{r^3}}{1 - \frac{2M}{r}} + \frac{\ddot N}{\left( 1 - \frac{2M}{r} \right)^2} 
 - \frac{\frac{M}{r^2} \dot N}{ \left( 1 - \frac{2M}{r} \right)^3} \, .
\end{align}
and therefore Eq.~(\ref{geodesicdev1B}) gives 
\begin{align}
\label{geodesicdev1BB0}
\frac{D^2 S^r}{d{\tilde\tau}^2} 
= \left\{ \frac{2M}{r^3} + \frac{\ddot N}{1 - \frac{2M}{r}} 
 - \frac{\frac{M}{r^2} \dot N}{ \left( 1 - \frac{2M}{r} \right)^2} \right\} \e^{-2N} S^r \, .
\end{align}
Furthermore, by using (\ref{lr2}), we rewrite (\ref{geodesicdev1BB0}) as follows, 
\begin{align}
\label{geodesicdev1BB}
\frac{D^2 S^r}{d{\tilde\tau}^2} 
= \left\{ \frac{2M \e^{-2N}}{r^3} 
+ \frac{\frac{dH}{d\tau} + H^2}{1 - \frac{2M}{r}} 
 - \frac{M H \e^{-N}}{ r^2 \left( 1 - \frac{2M}{r} \right)^2} \right\} S^r \, .
\end{align}
We now consider the meanings of each term in (\ref{geodesicdev1BB}). 

The first term corresponds to the tidal force induced by the black hole as usual. 
In the case of Type I or Big Rip singularity, $N$ diverges at the singularity and therefore the factor $\e^{-2N}$ in (\ref{geodesicdev1BB}) vanishes. 
This tells that the tidal force is strongly suppressed near the Type I singularity. 
This is merely because the surface gravity $\mathcal{K}$ in (\ref{sg3}) becomes smaller by the expansion of the universe due to the growth of the black hole. 
On the other hand, in the cases of Type II, III, and IV singularities, $N$ is finite, and there is no such suppression even near the singularities, although it becomes smaller 
by the expansion of the universe. 

The second term generates the standard future singularities in the region $r\to \infty$ as in (\ref{geodesicdev2}). 
In the case of the Type I and III future singularities, any extended object is torn by the tidal force. 
Due to the factor $\left( 1 - \frac{2M}{r} \right)^{-1}$, the tidal force is enhanced near the horizon 
as occurred in the case of the angular direction (\ref{geodesicdev1thph2}). 

The third term could be a new type of term, which is most dominant near the horizon due to the factor $\left( 1 - \frac{2M}{r} \right)^{-2}$. 
Note that the signature of the third term is negative, which shows that any extended object is not torn but pressed and crushed, as in the Big Crunch. 

For the spherical surface given by $r=r(t)$, which satisfies the following condition, 
\begin{align}
\label{nosurface}
0 = \frac{2M \e^{-2N}}{r^3} 
+ \frac{\frac{dH}{d\tau} + H^2}{1 - \frac{2M}{r}} 
 - \frac{M H \e^{-N}}{ r^2 \left( 1 - \frac{2M}{r} \right)^2} \, ,
\end{align}
we find that the tidal force vanishes on the surface. 
In the region $r>r(\tau)$, extended objects will be torn in Type I, II, and III singularities, but in the region $2M<r<r(\tau)$, these objects are pressed. 
In the case of the Type I, II, and III singularities, the second term in (\ref{geodesicdev1BB}) dominates near the singularity. 
In order to cancel the corresponding second term in (\ref{nosurface}), the last third term must become large so that $r\to 2M$. 
Therefore, near the future Type I, II, and III singularities, we find $r(t)\to 2M$. 

We may also consider the Little Rip scenario~\cite{Frampton:2011sp, Brevik:2011mm, Frampton:2011rh} when there is a black hole in the spacetime. 
In the Little Rip scenario, $H$ becomes too large, but $\frac{dH}{d\tau}$ is not always large, although $\e^N$ becomes large very rapidly. 

As an example, we may consider the case that 
\begin{align}
\label{lr1}
H = H_\mathrm{lr} \tau\, .
\end{align}
Here, $H_\mathrm{lr}$ is a constant. 
Then one gets
\begin{align}
\label{lr3}
\frac{dH}{d\tau}=H_\mathrm{lr}\, , \quad \e^N=\e^{\frac{H_\mathrm{lr} \tau^2}{2} + N_0}\, ,
\end{align}
with a constant $N_0$. 
In the late-time or far future, the first and the third terms in (\ref{geodesicdev1BB}) can be neglected, and we obtain, 
\begin{align}
\label{geodesicdev1lr}
\frac{D^2 S^r}{d{\tilde\tau}^2} 
\sim \frac{{H_\mathrm{lr}}^2 \tau^2}{1 - \frac{2M}{r}} S^r \, .
\end{align}
Therefore, the tidal force becomes larger and larger in the far future, and any extended objects are torn. 
The tidal force is enhanced near the horizon. 

In summary of this subsection, 
\begin{itemize}
\item In the case of Type I singularity, the singularity occurs, and any extended object is torn and ripped in the angular directions. 
The tidal force generating the rip is enhanced near the horizon when a black hole exists. 
Near the horizon, there is a region where the object is pressed in the radial direction, but the volume of the region vanishes at the singularity. 
Outside the region, any extended object is torn. 
The expansion of the universe suppresses the tidal force generated by the gravitational force of the black hole, and the tidal force vanishes at the singularity. 
\item In the case of Type II singularity, any extended object is pulled in the angular directions, but the kinematical work given by the tidal force could be finite, and 
therefore if the absolute value of the binding energy for the object is larger than the kinetic work, the object might not be ripped. 
The tidal force generating the rip is enhanced near the horizon when a black hole exists, but near the horizon, there is a region where the object is pressed 
in the radial direction. 
The volume of the region remains finite at the singularity. 
Outside the region, any extended object is pulled. 
The expansion suppresses the tidal force generated by the gravitational force of the black hole, but the tidal force does not vanish at the singularity, although it goes to zero 
by the expansion of the universe after the singularity. 
\item In the case of Type III singularity, any extended object is torn and ripped in the angular directions. 
The tidal force generating the rip is enhanced near the horizon when a black hole exists. 
Near the horizon, there is a region where the object is pressed in the radial direction, but the volume of the region vanishes at the singularity. 
Outside the region, any extended object is torn. 
The expansion suppresses the tidal force generated by the gravitational force of the black hole, but the tidal force does not vanish at the singularity. 
\item In the case of Type IV singularity, any extended object is pulled in the angular directions, but the kinematical work given by the tidal force could be finite, and 
therefore if the absolute value of the binding energy for the object is larger than the kinetic work, the object might not be ripped. 
The tidal force generating the rip is enhanced near the horizon when a black hole exists, but near the horizon, there is a region where the object is pressed 
in the radial direction. 
The volume of the region remains finite at the singularity. 
Outside the region, any extended object is pulled. 
The expansion suppresses the tidal force generated by the gravitational force of the black hole, but the tidal force does not vanish at the singularity, although it goes to zero 
by the expansion of the universe after the singularity. 
\item In the case of the Little Rip, although there is no finite future singularity, any extended object is finally torn and ripped in the angular directions. 
The tidal force generating the rip is enhanced near the horizon when a black hole exists, but near the horizon, there is a region where the object is pressed 
in the radial direction. 
The volume of the region goes to vanish by the expansion of the universe. 
Outside the region, any extended object is torn and ripped, finally even in the radial direction. 
The expansion suppresses the tidal force generated by the gravitational force of the black hole, but the tidal force goes to zero 
by the expansion of the universe after the singularity. 
\end{itemize}
For the observer in the universe without any black hole, the tidal force received by the observer is isotropic, that is, the tidal force does not depend on the direction. 
On the other hand, the observer near the black hole horizon could observe anisotropies of the tidal force, which could be enhanced near the horizon. 
Furthermore, the tidal force observed very near the black hole's horizon, the tidal force in the angular directions acts to tear any extended object, but the 
tidal force in the radial direction acts to press the object.

\section{New type of future singularity}\label{SecIV}

In this section, we consider a new type of future singularity. 
Cosmic censorship usually requires that the spacetime singularity should be hidden inside the horizon. 
The future singularities, including the Big Crunch, are naked singularities. 
We now consider the singularity on the sphere. 
When the radius of the sphere vanishes, the singularity lies at the origin, as in the standard black holes. 
Even if the radius becomes larger, as long as the radius is smaller than the radius of the horizon, the singularity is hidden and the cosmic censorship is not broken. 
If the radius of the singularity sphere becomes larger than the horizon radius, the singularity becomes naked. 
Furthermore, when the radius becomes infinite, the universe could end as in the future singularity case, although the singularity is not inhomogeneous. 

First, we show that the spacetime in (\ref{dsph}) can be transformed into the Schwarzschild-type spacetime as follows \cite{Nojiri:2020blr},
\begin{align}
\label{GBiv0}
ds^2 = - \e^{2\nu (\tilde r,\tilde t)} d{\tilde t}^2 + \e^{2\lambda (\tilde r,\tilde t)} d{\tilde r}^2 + {\tilde r}^2 \left( d\theta^2 + \sin^2\theta d\phi^2 \right)\, .
\end{align}
For the metric given by (\ref{dsph}), we redefine the radial coordinate $\tilde r$ using the following form,
\begin{align}
\label{G2}
{\tilde r}^2 \equiv D ( t , r ) \, ,
\end{align}
provided that $D ( t , r )>0$. 
Generally, Eq.~(\ref{G2}) is solvable with respect to $ r $, i.e., $ r = r ( t , \tilde r)$.
Subsequently, Eq.~(\ref{dsph}) can be reformulated in the following form,
\begin{align}
\label{G3}
ds^2 =&\, \left\{ - A \left( t , r \left( t , \tilde r\right) \right)
+ 2 B \left( t , r \left( t , \tilde r\right) \right) \frac{\partial r }{\partial t }
+ C \left( t , r \left( t , \tilde r\right) \right) \left( \frac{\partial r }{\partial t } \right)^2
\right\} d t ^2 \nonumber \\
&\, + 2 \left( B \left( t , r \left( t , \tilde r\right) \right)
+ C \left( t , r \left( t , \tilde r\right) \right) \frac{\partial r }{\partial t } \right)
\frac{\partial  r }{\partial \tilde r} d t  d\tilde r \nonumber \\
&\, + C \left( t , r \left( t , \tilde r\right) \right)
\left( \frac{\partial  r }{\partial \tilde r} \right)^2 d{\tilde r}^2
+ {\tilde r}^2 \left( d\theta^2 + \sin^2\theta d\phi^2 \right) \, .
\end{align}
Additionally, we represent $ t $ as $ t (t,r)$.
Hence, Eq. ~(\ref{G3}) can be expressed as,
\begin{align}
\label{G4}
ds^2 =&\, \left\{ - A \left( t \left(\tilde t, \tilde r\right),
 r \left( t \left(\tilde t, \tilde r\right), \tilde r\right) \right)
+ 2 B \left( t \left(\tilde t, \tilde r\right), r \left( t , \tilde r\right)
\right) \frac{\partial r \left( t , \tilde r\right)}{\partial t }
+ C \left( t , r \left( t , \tilde r\right) \right) \left( \frac{\partial r \left( t , \tilde r\right)}{\partial t } \right)^2
\right\} \nonumber \\
&\, \quad \times \left( \frac{\partial  t \left(\tilde t, \tilde r\right)}{\partial \tilde t} \right)^2 d{\tilde t}^2 \nonumber \\
&\, + 2 \left[ \left( B \left( t \left(\tilde t, \tilde r\right), r \left( t \left(\tilde t, \tilde r\right), \tilde r \right) \right)
+ C \left( t , r \left( t , \tilde r\right) \right) \frac{\partial  r \left( t , \tilde r\right)}{\partial  t } \right)
\frac{\partial  r \left( t , \tilde r\right)}{\partial \tilde r}
\frac{\partial  t \left(\tilde t, \tilde r\right)}{\partial \tilde t} \right. \nonumber \\
&\, \qquad + \left\{ - A \left( t \left(\tilde t, \tilde r\right),
 r \left( t \left(\tilde t, \tilde r\right), \tilde r\right) \right)
+ 2 B \left( t \left(\tilde t, \tilde r\right), r \left( t \left(\tilde t, \tilde r\right), \tilde r\right)
\right) \frac{\partial r \left( t , \tilde r\right)}{\partial t } \right. \nonumber \\
&\, \left. \left. \qquad \qquad + C \left( t , r \left( t , \tilde r\right) \right) \left( \frac{\partial r \left( t , \tilde r\right)}{\partial t } \right)^2 \right\}
\frac{\partial  t \left(\tilde t, \tilde r\right)}{\partial \tilde t}
\frac{\partial  t \left(\tilde t, \tilde r\right)}{\partial \tilde r} \right] d\tilde t d\tilde r \nonumber \\
&\, + \left[ C \left( t , r \left( t , \tilde r\right) \right)
\left( \frac{\partial  r \left( t , \tilde r\right)}{\partial \tilde r} \right)^2
+ 2 \left( B \left( t , r \left( t \left(\tilde t, \tilde r\right), \tilde r\right) \right)
+ C \left( t , r \left( t \left(\tilde t, \tilde r\right), \tilde r\right) \right) \frac{\partial r \left( t , \tilde r\right)}{\partial t } \right)
\frac{\partial  r \left( t , \tilde r\right)}{\partial \tilde r}
\frac{\partial  t \left(\tilde t, \tilde r\right)}{\partial \tilde r} \right. \nonumber \\
&\, \qquad + \left\{ - A \left( t \left(\tilde t, \tilde r\right),
 r \left( t \left(\tilde t, \tilde r\right), \tilde r\right) \right)
+ 2 B \left( t \left(\tilde t, \tilde r\right), r \left( t \left(\tilde t, \tilde r\right), \tilde r\right)
\right) \frac{\partial r \left( t , \tilde r\right)}{\partial t } \right. \nonumber \\
&\, \left. \left. \qquad \qquad + C \left( t , r \left( t , \tilde r\right) \right) \left( \frac{\partial r \left( t , \tilde r\right)}{\partial t } \right)^2
\right\} \left( \frac{\partial  t \left(\tilde t, \tilde r\right)}{\partial \tilde r} \right)^2
\right] d{\tilde r}^2 \nonumber \\
&\, + {\tilde r}^2 \left( d\theta^2 + \sin^2\theta d\phi^2 \right) \, .
\end{align}
We can choose the time-coordinate $\tilde t$ so that,
\begin{align}
\label{G5}
0=&\, \left( B \left( t \left(\tilde t, \tilde r\right), r \left( t \left(\tilde t, \tilde r\right), \tilde r\right) \right)
+ C \left( t , r \left( t , \tilde r\right) \right) \frac{\partial  r \left( t , \tilde r\right)}{\partial  t } \right)
\frac{\partial  r \left( t , \tilde r\right)}{\partial \tilde r}
\frac{\partial  t \left(\tilde t, \tilde r\right)}{\partial \tilde t} \nonumber \\
&\, + \left\{ - A \left( t \left(\tilde t, \tilde r\right),
 r \left( t \left(\tilde t, \tilde r\right), \tilde r\right) \right)
+ 2 B \left( t \left(\tilde t, \tilde r\right), r \left( t \left(\tilde t, \tilde r\right), \tilde r\right)
\right) \frac{\partial r \left( t , \tilde r\right)}{\partial t }
+ C \left( t , r \left( t , \tilde r\right) \right) \left( \frac{\partial r \left( t , \tilde r\right)}{\partial t } \right)^2
\right\} \nonumber \\
&\, \qquad \times \frac{\partial  t \left(\tilde t, \tilde r\right)}{\partial \tilde t}
\frac{\partial  t \left(\tilde t, \tilde r\right)}{\partial \tilde r} \, .
\end{align}
Through the above information, $\nu(\tilde t, \tilde r)$ and $\lambda(\tilde t, \tilde r)$ can yield the following forms,
\begin{align}
\label{GBiv}
 - \e^{2\nu (\tilde r, \tilde t)} \equiv &\, \left\{ - A \left( t \left(\tilde t, \tilde r\right),  r \left( t \left(\tilde t, \tilde r\right), \tilde r\right) \right)
+ 2 B \left( t \left(\tilde t, \tilde r\right), r \left( t \left(\tilde t, \tilde r\right), \tilde r\right) \right)
\left. \frac{\partial r \left( t , \tilde r\right)}{\partial t } \right|_{ t = t \left(\tilde t, \tilde r\right)} \right. \nonumber \\
&\, \left. \qquad + C \left(  t , r \left( t , \tilde r\right) \right) \left( \left. \frac{\partial r \left( t , \tilde r\right)}{\partial t } 
\right|_{ t = t \left(\tilde t, \tilde r\right)} \right)^2
\right\} \left( \frac{\partial  t \left(\tilde t, \tilde r\right)}{\partial \tilde t} \right)^2 \, , \nonumber \\
\e^{2\lambda (\tilde r, \tilde t)} \equiv\, & C \left( t , r \left( t , \tilde r\right) \right)
\left( \left. \frac{\partial  r \left( t , \tilde r\right)}{\partial \tilde r} \right|_{ t = t \left(\tilde t, \tilde r\right)} \right)^2 \nonumber \\
&\, + 2 \left( B \left( t , r \left( t \left(\tilde t, \tilde r\right), \tilde r\right) \right)
+ C \left( t , r \left( t \left(\tilde t, \tilde r\right), \tilde r\right) \right) \left. \frac{\partial r \left( t , \tilde r\right)}{\partial t }
\right|_{ t = t \left(\tilde t, \tilde r\right)} \right)
\left. \frac{\partial  r \left( t , \tilde r\right)}{\partial \tilde r} \right|_{ t = t \left(\tilde t, \tilde r\right)} 
\frac{\partial  t \left(\tilde t, \tilde r\right)}{\partial \tilde r} \nonumber \\
&\, + \left\{ - A \left( t \left(\tilde t, \tilde r\right),  r \left( t \left(\tilde t, \tilde r\right), \tilde r\right) \right)
+ 2 B \left( t \left(\tilde t, \tilde r\right), r \left( t \left(\tilde t, \tilde r\right), \tilde r\right)
\right) \left. \frac{\partial r \left( t , \tilde r\right)}{\partial t } \right|_{ t = t \left(\tilde t, \tilde r\right)} \right. \nonumber \\
&\, \left. \qquad \qquad + C \left( t , r \left( t , \tilde r\right) \right) \left( \left. \frac{\partial r \left( t , \tilde r\right)}{\partial t } 
\right|_{ t = t \left(\tilde t, \tilde r\right)} \right)^2 
\right\} \left( \frac{\partial  t \left(\tilde t, \tilde r\right)}{\partial \tilde r} \right)^2 \, .
\end{align}
Thus, we have shown that the spherically symmetric and time-dependent spacetime in (\ref{dsph}) can be rewritten in the form as in 
the Schwarzschild spacetime in (\ref{GBiv0}). 

For the FLRW universe (\ref{FLRW}), which can be rewritten as, 
\begin{align}
\label{FLRW2}
ds^2 = - d t ^2 + a( t )^2 \left\{ d r ^2 +  r ^2 \left( d\theta^2 + \sin^2\theta d\phi^2 \right) \right\}\, ,
\end{align}
we find
\begin{align}
\label{FLRW_ABCD}
A =1\, , \quad B=0 \, , \quad C=a( t )^2 \, , \quad
{\tilde r}^2 = D = a( t )^2  r ^2 \, .
\end{align}
Then Eq.~(\ref{G5}) reduces
\begin{align}
\label{G5B}
0 = - \frac{a'( t ) \tilde r}{a( t )} + \left( -1 + \frac{a'( t )^2 {\tilde r}^2}{a( t )^2} \right)
\frac{\partial  t \left(\tilde t, \tilde r\right)}{\partial \tilde r} \, .
\end{align}
Here we have used $ r =\frac{\tilde r}{a( t )}$, which is given by the last equation in (\ref{FLRW_ABCD}).
Eq.~(\ref{GBiv}) also gives
\begin{align}
\label{GBivBB}
 - \e^{2\nu (\tilde r, \tilde t)} =&\, \left( -1 + \frac{a'( t )^2 {\tilde r}^2}{a( t )^2} \right)
\left( \frac{\partial  t \left(\tilde t, \tilde r\right)}{\partial \tilde t} \right)^2 \, , \nonumber \\
\e^{2\lambda (\tilde r, \tilde t)} =&\, 1 { - \frac{2\tilde r a'( t )}{a( t )} \frac{\partial  t \left(\tilde t, \tilde r\right)}{\partial \tilde r} }
+ \left( -1 + \frac{a'( t )^2 {\tilde r}^2}{a( t )^2} \right) \left( \frac{\partial  t \left(\tilde t, \tilde r\right)}{\partial \tilde r} \right)^2 \, .
\end{align}

Metric~(\ref{GBiv0}) yields the following non-vanishing connections,
\begin{align}
\label{GBv}
&\Gamma^{\tilde t}_{\tilde t \tilde t}=\dot\nu \, , \quad \Gamma^{\tilde r}_{\tilde t \tilde t} 
= \e^{-2(\lambda - \nu)}\nu' \, , \quad \Gamma^{\tilde t}_{\tilde t \tilde r}=\Gamma^{\tilde t}_{\tilde r \tilde t}=\nu'\, , \quad
\Gamma^{\tilde t}_{\tilde r \tilde r} = \e^{2\lambda - 2\nu}\dot\lambda \, , \quad \Gamma^{\tilde r}_{\tilde t \tilde r} = \Gamma^{\tilde r}_{\tilde r\tilde t} 
= \dot\lambda \, , \nonumber \\
& \Gamma^{\tilde r}_{\tilde r\tilde r}=\lambda'\, , \quad \Gamma^i_{jk} = \bar{\Gamma}^i_{jk}\, ,\quad \Gamma^{\tilde r}_{ij}=-\e^{-2\lambda}\tilde r \bar{g}_{ij} \, , \quad
\Gamma^i_{\tilde rj}=\Gamma^i_{j\tilde r}=\frac{1}{\tilde r} \, \delta^i_{\ j}\,,
\end{align}
where the metric $\bar{g}_{ij}$ is given by
$\sum_{i,j=1}^2 \bar{g}_{ij} dx^i dx^j = d\theta^2 + \sin^2\theta d\phi^2$,
with $\left(x^1=\theta,\, x^2=\phi\right)$, and 
$\bar{ \Gamma}^i_{jk}$ represents the metric connection of $\bar{g}_{ij}$.
The ``dot'' and ``prime'' symbols denote differentiation with respect to $\tilde t$ and $\tilde r$, respectively.


Non-vanishing components of the curvatures are given by, 
\begin{align}
\label{curvaturesB}
R_{\tilde r \tilde t \tilde r \tilde t} = & - \e^{2\lambda} \left\{ \ddot\lambda + \left( \dot\lambda - \dot\nu \right) \dot\lambda \right\}
+ \e^{2\nu}\left\{ \nu'' + \left(\nu' - \lambda'\right)\nu' \right\} \, ,\nonumber \\
R_{\tilde t i \tilde t j} =& \, \tilde r\nu' \e^{-2\lambda + 2\nu} \bar{g}_{ij} \, ,\nonumber \\
R_{\tilde r i \tilde r j} =& \, \lambda' \tilde r \bar{ g}_{ij} \, ,\quad 
{R_{\tilde t i \tilde r j}= \dot\lambda \tilde r \bar{ g}_{ij} } \, , \quad
R_{ijkl} = \left( 1 - \e^{-2\lambda}\right) {\tilde r}^2 \left(\bar{g}_{ik} \bar{g}_{jl} - \bar{g}_{il} \bar{g}_{jk} \right)\, ,\nonumber \\
R_{\tilde t \tilde t} =& - \left\{ \ddot\lambda + \left( \dot\lambda - \dot\nu \right) \dot\lambda \right\}
+ \e^{-2\lambda + 2\nu} \left\{ \nu'' + \left(\nu' - \lambda'\right)\nu' + \frac{2\nu'}{\tilde r}\right\} \, ,\nonumber \\
R_{\tilde r \tilde r} =& \, \e^{2\lambda - 2\nu} \left\{ \ddot\lambda + \left( \dot\lambda - \dot\nu \right) \dot\lambda \right\}
 - \left\{ \nu'' + \left(\nu' - \lambda'\right)\nu' \right\}
+ \frac{2 \lambda'}{\tilde r} \, ,\quad
R_{\tilde t \tilde r} =R_{\tilde r \tilde t} = \frac{2\dot\lambda}{\tilde r} \, , \nonumber \\
R_{ij} =&\, \left\{ 1 + \left\{ - 1 - \tilde r \left(\nu' - \lambda' \right)\right\} \e^{-2\lambda}\right\} \bar{g}_{ij}\ , \nonumber \\
R=& \, 2 \e^{-2 \nu} \left\{ \ddot\lambda + \left( \dot\lambda - \dot\nu \right) \dot\lambda \right\} + \e^{-2\lambda}\left\{ 
 - 2\nu'' - 2\left(\nu' - \lambda'\right)\nu' - \frac{4\left(\nu' - \lambda'\right)}{\tilde r} + \frac{2\e^{2\lambda} - 2}{{\tilde r}^2} \right\} \, ,
\end{align}
Then the Einstein equation (\ref{Ee}) has the following forms, 
\begin{align}
\label{scalr}
\e^{-2\lambda} \left( \frac{2\lambda'}{\tilde r} - \frac{1}{{\tilde r}^2} \right) + \frac{1}{r^2} =&\, \kappa^2 \rho \, , \nonumber \\
\e^{- 2\lambda} \left( \frac{2\nu'}{\tilde r} + \frac{1}{{\tilde r}^2} \right) - \frac{1}{{\tilde r}^2} =&\, \kappa^2 p_\mathrm{rad} \, , \nonumber \\
 - \e^{- 2\lambda} \left( \nu'' + {\nu'}^2 - \nu'\lambda' + \frac{\nu' - \lambda'}{\tilde r} \right) 
 - \left( \dot\nu \dot\lambda -\ddot\lambda - \dot\lambda^2 \right) \e^{-2\nu}=&\, \kappa^2 p_\mathrm{ang} \, , \nonumber \\
\frac{2\dot\lambda}{\tilde r}=&\, \kappa^2 j \, .
\end{align}
By deleting $\tilde t$ and $\tilde r$, we may obtain equations of state. 

As an example of the scenario for the new singularity, we consider the following, 
\begin{align}
\label{ex1}
\e^{2\nu}=\left( 1 - \frac{r_0}{\tilde r} \right) \left( 1 - \frac{\alpha\left(\tilde t\right)}{\tilde r} \right) \, , \quad 
\e^{-2\lambda}=\left( 1 - \frac{r_0}{\tilde r} \right) \, .
\end{align}
Here $\alpha\left(\tilde t\right)$ is a function of $\tilde t$, which gives the time-evolution of the spacetime. 
The form of  $\alpha\left(\tilde t\right)$ is more specified later. 

\begin{figure}
\begin{center}

\scalebox{0.4}{
\unitlength=2mm
\begin{picture}(200,100)

\put(100,25){\makebox(0,0){\LARGE white hole}}
\put(100,63){\makebox(0,0){\LARGE black hole}}
\put(100,75){\makebox(0,0){\LARGE black hole singularity}}
\put(140,75){\makebox(0,0){\LARGE future singularity}}
\put(181,63){\makebox(0,0){\LARGE future null inifinity}}
\put(181,60){\makebox(0,0){\LARGE in our universe}}
\put(17,63){\makebox(0,0){\LARGE future null inifinity}}
\put(17,60){\makebox(0,0){\LARGE in another universe}}
\put(170,26){\makebox(0,0){\LARGE past null inifinity}}
\put(170,23){\makebox(0,0){\LARGE in our universe}}
\put(30,26){\makebox(0,0){\LARGE past null inifinity}}
\put(30,23){\makebox(0,0){\LARGE in another universe}}

\thicklines

\put(20,50){\line(1,1){20}}
\put(20,50){\line(1,-1){40}}
\put(60,10){\line(1,1){60}}
\put(80,70){\line(1,-1){60}}
\put(140,10){\line(1,1){40}}
\put(160,70){\line(1,-1){20}}

\put(60,10){\line(1,-2){1}}
\put(139,12){\line(1,-2){1}}
\multiput(61,8)(4,0){20}{\line(1,2){2}}
\multiput(63,12)(4,0){19}{\line(1,-2){2}}

\put(40,70){\line(1,-2){1}}
\put(159,72){\line(1,-2){1}}
\multiput(41,68)(4,0){30}{\line(1,2){2}}
\multiput(43,72)(4,0){29}{\line(1,-2){2}}

\end{picture}
}

\end{center}
\caption{The Penrose diagram of a new type of singularity in (\ref{ex1}). 
The black hole singularity extends outside of the horizon and becomes naked. 
The singularity turns out to be a finite future singularity. }
\label{Fig1}

\end{figure}
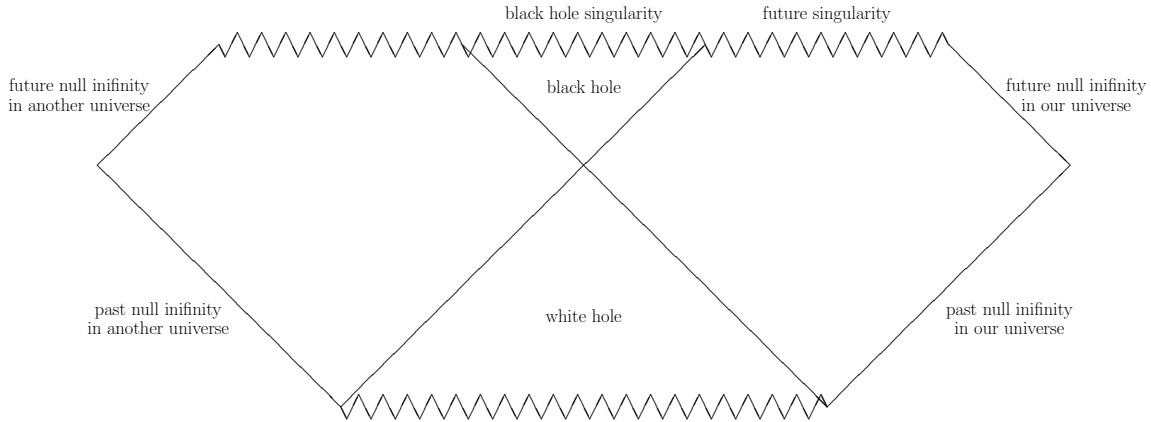

Because 
\begin{align}
\label{nlmbd}
\nu'=&\, \frac{1}{2} \left\{ \frac{\frac{r_0}{{\tilde r}^2}}{1 - \frac{r_0}{\tilde r}} + \frac{\frac{\alpha\left( \tilde t \right)}{{\tilde r}^2}}{1 - \frac{\alpha\left(\tilde t\right)}{\tilde r}} \right\} \, , \quad 
\lambda'= - \frac{1}{2} \frac{\frac{r_0}{{\tilde r}^2}}{1 - \frac{r_0}{\tilde r}} \, , \nonumber \\
\nu''=&\, - \frac{1}{2} \left\{ \frac{ \frac{2r_0}{{\tilde r}^3}}{1 - \frac{r_0}{\tilde r}} + \frac{\frac{{r_0}^2}{{\tilde r}^4}}{\left(1 - \frac{r_0}{\tilde r}\right)^2} 
+ \frac{\frac{2\alpha\left(\tilde t\right)}{{\tilde r}^3}}{1 - \frac{\alpha\left(\tilde t\right)}{{\tilde r}}} + \frac{\frac{\alpha\left(\tilde t\right)^2}{{\tilde r}^4}}{\left(1 - \frac{\alpha\left(\tilde t\right)}{\tilde r}\right)^2} \right\} \, ,
\end{align}
we obtain the following expressions of the curvatures, 
\begin{align}
\label{exR}
R_{\tilde t \tilde t}=&\, \left( 1 - \frac{r_0}{\tilde r} \right)^2 \left( 1 - \frac{\alpha\left(\tilde t\right)}{\tilde r} \right) \left\{
\frac{\frac{3r_0\alpha\left(\tilde t\right)}{{\tilde r}^4}}{4\left(1 - \frac{r_0}{\tilde r}\right)\left(1 - \frac{\alpha\left(\tilde t\right)}{\tilde r}\right)}
 - \frac{\frac{\alpha\left(\tilde t\right)^2}{{\tilde r}^4}}{4\left(1 - \frac{\alpha\left(\tilde t\right)}{\tilde r}\right)^2} \right\} \nonumber \\
R_{\tilde r \tilde r}=&\, - \frac{\frac{3r_0\alpha\left(\tilde t\right)}{{\tilde r}^4}}{4\left(1 - \frac{r_0}{\tilde r}\right)\left(1 - \frac{\alpha\left(\tilde t\right)}{\tilde r}\right)}
+ \frac{\frac{\alpha\left(\tilde t\right)}{{\tilde r}^3}}{1 - \frac{\alpha\left(\tilde t\right)}{\tilde r}} + \frac{\frac{\alpha\left(\tilde t\right)^2}{{\tilde r}^4}}{4\left(1 - \frac{\alpha\left(\tilde t\right)}{\tilde r}\right)^2} \nonumber \\
R_{\tilde t \tilde r}=&\, R_{\tilde r \tilde t} = 0 \, , \nonumber \\
R_{ij}=&\, \left[ 1 + \left( 1 - \frac{r_0}{\tilde r} \right) \left\{ - 1 - \frac{1}{2} \left( \frac{\frac{2r_0}{\tilde r}}{1 - \frac{r_0}{\tilde r}} 
+ \frac{\frac{\alpha\left(\tilde t\right)}{\tilde r}}{1 - \frac{\alpha\left(\tilde t\right)}{\tilde r}} \right)
\right\} \right] \, , \nonumber \\
R =&\, 2 \left( 1 - \frac{r_0}{\tilde r} \right) \left\{
\frac{\frac{3r_0\alpha\left(\tilde t\right)}{{\tilde r}^4}}{4\left(1 - \frac{r_0}{\tilde r}\right)\left(1 - \frac{\alpha\left(\tilde t\right)}{\tilde r}\right)}
 - \frac{\frac{\alpha\left(\tilde t\right)}{{\tilde r}^3}}{1 - \frac{\alpha\left(\tilde t\right)}{\tilde r}} 
 - \frac{\frac{\alpha\left(\tilde t\right)^2}{{\tilde r}^4}}{4\left(1 - \frac{\alpha\left(\tilde t\right)}{\tilde r}\right)^2} \right\} \nonumber \\
\end{align}
There is no singularity at the horizon $\tilde r=r_0$, but there is one at $\tilde r=\alpha\left(\tilde t\right)$. 
When $r_0>\alpha\left(\tilde t\right)$, the singularity is hidden by the horizon, but the singularity becomes naked and if 
$\alpha\left(\tilde t\right)\to \infty$ in the finite future, as in the Big Rip singularity, the end of the universe appears. 
We should also note that the last equation in (\ref{scalr}) with (\ref{ex1}) tells $j=0$, that is, there is no accretion of matter, 
which is different from the examples in the previous section. 
The Penrose diagram of the spacetime is depicted in FIG.~\ref{Fig1}. 

The general Penrose diagram expresses the spherically symmetric spacetime. 
The horizontal direction is the spatial direction corresponding to the radius and the vertical direction is the temporal direction. 
Then the horizontal lines correspond to the spatial hypersurface and 
the null direction or light-like direction is expressed by the direction of $\pm 45$ degrees. 
Then, for example, the paths of light rays are given by the $\pm 45$ lines. 

The spacetime in FIG.~\ref{Fig1} could be realised by a fluid satisfying (\ref{scalr}), although the EoS might be complicated. 
Even if we do not use the unknown fluid, we may realise the spacetime by using two scalar fields as \cite{Nojiri:2020blr} and without ghost \cite{Nojiri:2023dvf}. 

\section{Summary and Discussions}\label{SecV}

This paper is devoted to studying the dynamical black hole in the background of the expanding universe. 
When the expansion of the corresponding universe is accelerating, the universe may approach a future singularity. 
We found such black hole solutions by using the cosmic fluid, especially when the EoS parameter of the fluid is a constant, as in (\ref{EoS}). 
It is shown that the Schwarzschild black hole (\ref{Schwrz}) embedded in the expanding universe is an exact solution of the Einstein equation. 
The exact solution is different from the McVittie solution (\ref{McV}) \cite{McVittie:1933zz}. 

For the solution under consideration, the black hole mass increases due to the expansion of the universe. 
Therefore, during the inflation, the primordial black holes become larger and larger and after the inflation, (super-)massive black holes appear. 
By using this scenario, we may explain the origin of the black holes observed in the GW231123 event, in addition to the dark matter, 
and also supermassive black holes at the centre of galaxies. 

We also considered general EoS (\ref{EoS4}) and several kinds of future singularities, the Type I, II, III, and IV, and also an oscillating universe 
(\ref{T9}) generated by the fluid governed by the general EoS (\ref{T13}). 
The thermodynamics of our expanding spacetime with a dynamical black hole is also studied in this paper. 
There are two heat sources corresponding to the black hole horizon and the cosmological horizons, so there could occur several kinds of transitions, 
as in the Hawking-Page phase transition \cite{Hawking:1982dh}. 
It is shown that the black hole horizon enhances the tidal force, but near the horizon, the tidal force works to press the extended object 
by using the geodesic deviation equation (\ref{geodesicdev}) \cite{Wald:1984rg} for the angular directions (\ref{geodesicdev1thph2}) and the radial direction (\ref{geodesicdev1BB0}). 

A new type of future singularity is proposed by considering a spherical singularity. 
The metric of the corresponding spacetime is given in (\ref{GBiv0}) with (\ref{ex1}). 
As depicted in the Penrose diagram in FIG.~\ref{Fig1}, 
the radius of the spherical singularity becomes larger than the black hole horizon radius, and it becomes naked. 
The universe may end up when the radius of the singularity becomes infinite. 

In Ref.~\cite{Katsuragawa:2024bwm}, future singularities originating from the anisotropies in the Universe have been investigated. 
The singularity may include that which is called the Big Twist, where the off-diagonal elements of the metric play the central role. 
The new class of singularities are investigated in the homogeneous and anisotropic universe, and they are compared with the known singularities 
in the homogeneous and isotropic universe. 
It could be interesting to investigate these anisotropic singularities for a black hole under investigation, especially the role and behaviour of the horizon. 


Finally, we consider the possibility that the shadow of the black hole described in this paper could be observed as in the Event Horizon Telescope~\cite{EventHorizonTelescope:2019dse}. 
For this purpose, we consider the geodesic motion of a particle, including the case of a photon. 
The Lagrangian of the particle in the spacetime described by the metric (\ref{geo2}) is given by, 
\begin{align}
\label{L}
\mathcal{L}=\frac{1}{2}\e^{2N(t)} \left\{ - \e^{2\nu(r)}{\overset{\circ}{t}}^2 + \e^{2\lambda(t)} {\overset{\circ}{r}}^2 
+r^2 \left( {\overset{\circ}{\theta}}^2 + \sin^2\theta {\overset{\circ}{\phi}}^2 \right) \right\}\, . 
\end{align}
Here $\overset{\circ}{t}$ is a derivative of $t$ with respect to the parameter $\lambda$, which parametrises the orbit of the particle, 
$\overset{\circ}{t}\equiv \frac{dt}{d\lambda}$, etc. 
When we consider a massless particle such as a photon, we have $\mathcal{L}=0$ and when we consider a particle with a mass $m$, $\mathcal{L}=-\frac{1}{2}m^2$. 
In the system, the angular momentum is conserved because the spacetime is spherically symmetric, although the energy is not conserved because the spacetime is time-dependent. 
When angular momentum is conserved, the orbit is confined to the plane which is perpendicular to the vector of the angular momentum. 
Therefore, without any loss of generality, we may choose $\theta=\frac{\pi}{2}$. 
Then the Lagrangian $L$ is reduced to the following form, 
\begin{align}
\label{L2}
\mathcal{L}=\frac{1}{2}\e^{2N(t)} \left( - \e^{2\nu(r)}{\overset{\circ}{t}}^2 + \e^{2\lambda(t)} {\overset{\circ}{r}}^2 
+r^2 {\overset{\circ}{\phi}}^2 \right) \, . 
\end{align}
Furthermore, the angular momentum $L$, which is now the momentum conjugate to $\phi$, is conserved, 
\begin{align}
\label{L3}
L\equiv \frac{\partial\mathcal{L}}{\overset{\circ}{\phi}} = \e^{2N(t)} r^2 \overset{\circ}{\phi}\, .
\end{align}
By choosing $\lambda=t$, we obtain $\overset{\circ}{t} = 1$, $\overset{\circ}{r} = \dot r$ and 
\begin{align}
\label{L4}
\frac{1}{2}{\dot r}^2 = \frac{1}{2} \left( \e^{2\nu(r) - 2\lambda(r)} - \frac{L^2\e^{-2N(t)}}{r^2} - m^2 \e^{-2N(t) - 2\lambda(r)} \right)\, .
\end{align}
In the massless case, $m=0$, the radial coordinate $r$ cannot be constant, which contradicts Eq.~(\ref{L4}). 
Therefore, there is no photon sphere, which is not inconsistent with the existence of the black hole shadow. 
Usually, the period that the photon circles the photon sphere is much smaller than the timescale of the expansion, and there appears approximately a photon sphere 
and the black hole shadow associated with the sphere.

It is also interesting that due to the dynamical behaviour of the black hole under consideration, its shadow may be different at different epochs.
Then there appears a natural conjecture that the black hole shadow may be consistent with observational bounds from EHT, at least at some epoch.
Even if it looks like the shadow is not consistent with EHT bounds for some gravity theory~\cite{Khodadi:2024ubi, Vagnozzi:2022moj, Nojiri:2024txy}, 
it may be perfectly O.K. at another epoch of the universe's evolution.

Note that the real radius is not given by $r$ but $\e^{N(t)}r$. 
Usually, the expansion of the universe makes the observed apparent radius of the black hole shadow smaller, but for the black hole of this paper, the horizon radius, and 
therefore, the radius of the black hole shadow grows proportional to $\e^{N(t)}$. 
Therefore, the observed radius of the black hole shadow does not decrease, but it could be constant.
Unfortunately, at present, we do not have any idea how to observe this phenomenon. 

We should also note that for a Type I or Big Rip singularity, $N(t)$ diverges at the singularity, which shows us that the second and third terms in the right-hand side of 
Eq.~(\ref{L4}) can be neglected. Hence, even for a massive particle, $m\neq 0$, the orbit of the particle goes to a light-like one. 
In the case of Types II, III, IV singularities, $N(t)$ is finite even at the singular point. 
This shows that the particles will go through the singularity, although in the case of the Type III singularity, any extended objects could be torn and ripped. 
Therefore, for the Type II and IV singularities, we may consider the universe after the singularity, that is, there might be singularities in the past.

One remark is in order. 
It has been considered a static black hole metric within a dark energy universe in Ref.~\cite{He:2017alg}. 
The account of the dynamics of the expanding universe, as in this paper, may lead to qualitative changes in the case of a static black hole. 
Furthermore, in Ref.~\cite{He:2017alg}, the lensing effect was used for the static black hole. 
It could be interesting to reconsider the lensing for the dynamical black holes in this paper, especially by including the massive primordial black holes.

\section*{Acknoweledgements} 

We are grateful to Professor Hong-Jian~He for the interest to this work.

\end{document}